\documentclass[acmsmall,nonacm]{acmart}

\AtBeginDocument{%
  }

\acmISBN{978-1-4503-XXXX-X/18/06}

\usepackage{cleveref}
\usepackage{subcaption}

\author{Vinay Koshy}
\email{vkoshy2@illinois.edu}
\affiliation{
    \institution{University of Illinois, Urbana-Champaign}
    \city{Urbana}
    \state{Illinois}
    \country{USA}
}
\author{Frederick Choi}
\email{fc20@illinois.edu}
\affiliation{
    \institution{University of Illinois, Urbana-Champaign}
    \city{Urbana}
    \state{Illinois}
    \country{USA}
}

\author{Yi-Shyuan Chiang}
\email{ysc6@illinois.edu}
\affiliation{
    \institution{University of Illinois, Urbana-Champaign}
    \city{Urbana}
    \state{Illinois}
    \country{USA}
}

\author{Hari Sundaram}
\email{hs1@illinois.edu}
\affiliation{
    \institution{University of Illinois, Urbana-Champaign}
    \city{Urbana}
    \state{Illinois}
    \country{USA}
}

\author{Eshwar Chandrasekharan}
\email{hs1@illinois.edu}
\affiliation{
    \institution{University of Illinois, Urbana-Champaign}
    \city{Urbana}
    \state{Illinois}
    \country{USA}
}

\author{Karrie Karahalios}
\email{kkarahal@illinois.edu}
\affiliation{
    \institution{University of Illinois, Urbana-Champaign}
    \city{Urbana}
    \state{Illinois}
    \country{USA}
}

\renewcommand{\shortauthors}{Vinay Koshy et al.}

\begin{document}

\title{Venire: A Machine Learning-Guided Panel Review System for Community Content Moderation}


\begin{abstract}
Research into community content moderation often assumes that moderation teams govern with a single, unified voice. However, recent work has found that moderators disagree with one another at modest, but concerning rates. The problem is not the root disagreements themselves. Subjectivity in moderation is unavoidable, and there are clear benefits to including diverse perspectives within a moderation team. Instead, the crux of the issue is that, due to resource constraints, moderation decisions end up being made by individual decision-makers. The result is decision-making that is inconsistent, which is frustrating for community members. To address this, we develop Venire, an ML-backed system for panel review on Reddit. Venire uses a machine learning model trained on log data to identify the cases where moderators are most likely to disagree. Venire fast-tracks these cases for multi-person review. Ideally, Venire allows moderators to surface and resolve disagreements that would have otherwise gone unnoticed. We conduct three studies through which we design and evaluate Venire: a set of formative interviews with moderators, technical evaluations on two datasets, and a think-aloud study in which moderators used Venire to make decisions on real moderation cases. Quantitatively, we demonstrate that Venire is able to improve decision consistency and surface latent disagreements. Qualitatively, we find that Venire helps moderators resolve difficult moderation cases more confidently. Venire represents a novel paradigm for human-AI content moderation, and shifts the conversation from replacing human decision-making to supporting it.\end{abstract}

\ccsdesc[500]{Human-centered computing~Collaborative and social computing systems and tools}
\ccsdesc[500]{Human-centered computing~Empirical studies in collaborative and social computing}
\ccsdesc[300]{Human-centered computing~Empirical studies in HCI}
\ccsdesc[300]{Human-centered computing~Collaborative interaction}
\keywords{Content moderation, human-AI interaction, decision-making, online communities}

\received{20 February 2007}
\received[revised]{12 March 2009}
\received[accepted]{5 June 2009}

\maketitle
\section{Introduction}\label{sec:introduction}

In many online communities, a team of volunteer moderators is responsible for creating and enforcing rules to govern acceptable speech. Although rules are often collectively set by the moderation team, moderators act more independently when enforcing them within the community. This implies a tension inherent to moderation efforts. While community policy reflects a moderation team's shared values, the day-to-day practice of this policy relies on individual decision makers. Prior work has found that moderators themselves disagree over how to apply community rules at concerning rates -- audits of moderator decision-making have found disagreement rates as high as 16\% \cite{koshy2023measuring} and 23\% \cite{lucas2019understanding}. This raises concerns that moderators on the same team may be enforcing rules inconsistently. Such inconsistency can be frustrating for community members. If users see that their post has been removed while similar posts remain unmoderated, they may feel unfairly singled out or targeted \cite{jhaver2019did, ma2022fairness}. Further, inconsistent rule application makes it harder for community members to learn a group's norms. Naively, one might try to increase decision consistency by requiring every moderation case be reviewed by a panel of moderators \cite{resnick2021survey}. In practice, this is infeasible. Universal panel review creates far too much extra work for moderation teams, which are already stretched thin \cite{li2022measuring, chandrasekharan2019crossmod}. In this paper, we explore an alternative approach. Instead of mandating universal panel review, we build Venire\footnote{The name is derived from "\textit{venire facias}", an archaic legal term for a written order from a judge directing a sheriff to assemble a jury \cite{wiki2022venire}}, a machine learning-guided panel review system.  

Venire acts as an overlay for Reddit's existing moderation queue, and gives moderators the ability to manually flag cases for panel review. When a case undergoes panel review, its final outcome is determined by a vote amongst multiple moderators rather than by a single decision-maker. Venire is backed by an ML model which uses moderation log data to predict how each team member would respond to an incoming case. Venire recommends a case for panel review when it predicts the moderation team is likely to disagree over how to handle it. Ideally, Venire helps moderation teams surface latent disagreements, while keeping the increase in moderator workload to a minimum. We present a series of three studies through which we design, build, and evaluate Venire. First, we conducted preliminary interviews to assess whether our intended goals for Venire were aligned with moderator needs.Second, we performed  two technical evaluations to ensure that Venire could deliver on the promising of increasing decision consistency and surfacing latent disagreements. Finally, we investigated how moderators use the system in practice by performing think-aloud study. In the think aloud study, moderators used Venire to make decisions on real moderation cases pulled from the r/ChangeMyView subreddit. 




Our initial interviews revealed that moderators were open to the idea of using a system like Venire. Moderators appreciated having a potential built-in channel as an alternative to informal deliberation processes already occurring within their communities. Moderators anticipated that the ML panel recommendations could help surface disagreements that were being missed. We also surfaced potential benefits we had not anticipated. Most critically, moderators felt Venire had value as an onboarding tool for new moderators. Because Venire provides a preliminary assessment of how contentious a case will be, it can avert controversial rulings from a new moderator, and increase a newbie moderator's confidence when handling straightforward cases.

In our technical evaluations, we conducted simulation-based analysis on two dataset to assess the quality of our ML model's panel recommendations. Using a large, publicly available toxicity dataset \cite{kumar2021designing}, we demonstrate that our predictive model can effectively triage moderation cases for panel review: we are able to approximate the decision-consistency benefits of universal panel review while assigning only a third cases of moderation cases to a panel. To make our analysis more ecologically valid, we constructed a smaller dataset containing labels from Prolific crowdworkers. To improve ecological validity, crowdworkers were asked to enforce an actual moderation rule from r/ChangeMyView, a popular Reddit community. Even on this smaller dataset, our model was able to anticipate controversial moderation cases about as accurately as a human rater. However, panel allocation was less efficient compared to in the toxicity dataset---it took assigning 60\% of all cases to a panel to approximate the decision consistency of universal panel review. Still, our results with both datasets indicate that model-assigned panels significantly outperformed random panel assignment at improving decision-consistency and surfacing disagreements.

Finally, we tested Venire's capabilities as an onboarding tool in our think-aloud study with moderators. We had moderators from other communities enforce a rule from r/ChangeMyView, using real data from r/ChangeMyView's mod queue. Moderators found Venire's ML model recommendations helpful for deciding when to assign a case for panel review, and noted that the panel review empowered them to play a more active role in the moderation queue. Taken together, our findings demonstrate how an ML-guided panel review system can support the reflective practice of moderation work \cite{cullen2022reflexive}, and facilitate high quality decision making. 
\section{Background}\label{sec:background}

The goal of this work is to develop a human-AI workflow for surfacing disagreements over subjective content moderation cases. As such, we present a review of prior work on: existing practices amongst community moderators for establishing a consistent moderation policy, prior attempts to build human-AI moderation tools, and machine learning approaches for modeling subjective decision-making. 

\subsection{Community Practices for Improving Consistency}\label{sec:background-practices}

Most commonly, moderators of online communities create sets of shared guidelines for rule enforcement to ensure the team acts in a consistent matter \cite{seering2019engagment, chandrasekharan2019crossmod, koshy2023measuring, dosono2019aapi, cullen2022reflexive, shahid2024whatsapp, seering2023twitch}. In interviews, both \citet{seering2019engagment} and \citet{cullen2022reflexive} find that these guidelines are developed iteratively over a community's lifespan, often in response to specific incidents where moderators felt a user crossed the line. Policy tends to be set collective by mod teams \cite{seering2019engagment, cullen2022reflexive, dosono2019aapi, seering2023twitch}, though moderators sometimes take community input, or defer to a ``head'' moderator \cite{seering2019engagment, cullen2022reflexive}. Occasionally, disputes over such policy can cause communities to fracture, splitting off into separate groups \cite{dosono2019aapi}. Even with an informal policy in place, moderators still consult with another when they are unsure how to handle a particular case \cite{seering2019engagment, cullen2022reflexive}. \citet{cullen2022reflexive} notes that such incidents can reveal places where moderators' mental models of how a rule should work, or even core values, are misaligned, creating an opportunity for reflection. 


Still, not every community's moderation team revolves around a comprehensive, iteratively developed policy. For example, \citet{seering2019engagment} found that at least one moderator was told to ``do whatever you feel makes the [community] better'' when they were made a moderator. Similarly, across studies, many interviewees report experiencing limited onboarding when they started as moderators, instead relying on more implicit processes to develop a feel for the community norms \cite{seering2019engagment, cullen2022reflexive, shahid2024whatsapp}. 

To our knowledge, only two studies have tried to directly assess how often moderators disagree with one another \cite{koshy2023measuring, lucas2019understanding}. The studies had moderators review sets of comments previously posted to two different large, discussion-based subreddits \cite{koshy2023measuring, lucas2019understanding}. For comments that received two in-study annotations, \citet{lucas2019understanding} found a disagreement rate of 23\% (Fleiss' $\kappa=0.46,N=222$), while \citet{koshy2023measuring} found a disagreement rate of 13\% ($N=134$). When comparing in-study labels to real life outcomes, the rates of disagreement were 28\% ($N=246$ annotations across 134 comments) and 26\% ($N=1020$ annotations across 798 comments) respectively. The present work is primarily motivated by the prevalence of content moderation disagreements found in these two studies, in spite of the existing measures that online communities take to ensure consistency.

\subsection{Human-AI Content Moderation Tools}\label{sec:background-tools}

Researchers in the CHI and CSCW communities have built a number of human-in-the-loop AI systems for content moderation \cite{huang2024opportunities}. These tools can be categorized as facilitating either \textit{top-down} \cite{choi2023convex, chandrasekharan2019crossmod, samory2021positive, hsieh2023nip, kuo2024wikibench, halfaker2020ores} or \textit{personalized} content moderation \cite{im2020synthesized, jhaver2023personalizing}. Although tools within each category tend to follow similar design patterns, they differ substantially in terms of the task the underlying machine learning model is trained on, the way model predictions are presented in an interface, and the place in the moderation process human decision-making is utilized. Because our work falls under the umbrella of top-down content moderation tools, we provide a more detailed review of prior top-down approaches. 

\subsubsection{AI tools for top-down moderation} Typically in \textit{top-down} tools, a machine learning model is used to flag content for a human moderator to review---moderator decisions in turn affect all users within a given community or platform \cite{choi2023convex, chandrasekharan2019crossmod, samory2021positive, hsieh2023nip, kuo2024wikibench, halfaker2020ores, gordon2022jury}. Amongst top-down tools, one of the key differentiating factors is the source of the training data. One approach is to use a generalized model. For example, in building a tool for Discord moderation, \citet{choi2023convex} use Perspective API.\footnote{https://www.perspectiveapi.com} The Perspective model is trained using crowdworker toxicity labels on comments across multiple social media platforms. In contrast, to build a tool for Reddit moderation, \citet{chandrasekharan2019crossmod} use historical data scraped from Reddit to train an ensemble of models that predict whether comments in specific communities will get removed or not. \citet{halfaker2020ores} adopt the most bespoke approach, creating a ``Wiki Labels'' system through which Wikipedians can contribute training data labels to specific model development requests.

Notably, almost all existing tools have been built with the goal of either reducing moderator labor, or helping moderators identify norm violations they would have otherwise missed \cite{choi2023convex, chandrasekharan2019crossmod, hsieh2023nip, halfaker2020ores}. Our approach, using machine learning models to improve the consistency of human decision-making, is relatively unique in this regard. 

Still, a few other researchers have also built tools that center disagreement in the content moderation process \cite{kuo2024wikibench, gordon2022jury}. In developing the jury learning framework, \citet{gordon2022jury} argue that rather than predicting a majority vote or average label across annotators, machine learning models should be trained to predict a label for each annotator in the dataset, using the annotators ID and demographic information as features. They create an interface on top of a model trained in this that allows the end user to choose which voices in the training dataset to amplify within their communities' automated tool. \citet{kuo2024wikibench}, whose work is perhaps most similar to our own, develop a tool that allows communities to curate evaluation datasets for AI tools they might want to adopt. As community members annotate data points, cases with disagreements are prioritized to receive additional ratings. Although our work shares a similar focus on contentious moderation cases, the goal of our work is to use a model to allocate panels efficiently, rather than to use panels to more accurately evaluate a model.

\subsection{Modeling Subjective Decision-Making}\label{sec:background-technical}

Gordon et al.'s jury learning framework \cite{gordon2022jury} is part of a broader trend amongst HCI and machine learning researchers, recognizing that modeling individual annotator beliefs can be beneficial for subjective tasks \cite{cabitza2023toward, rottger2021two, fleisig2023majority, fleisig2024perspectivist}. This viewpoint, sometimes referred to as ``perspectivism,'' is accompanied by varying motivations. Drawing on feminist theory, \citet{blackwell2017classification} argue that traditional majority vote aggregations of annotator labels can reinforce the viewpoints of dominant social groups. Other researchers appeal to more technical benefits of perspectivist approaches: that they more accurately represent the data generating process \cite{cabitza2023toward, fleisig2024perspectivist}, that they provide the ability to capture uncertainty in training labels that arise from human variation \cite{fleisig2024perspectivist, cabitza2023toward, gordon2022jury}, and that they afford end users more flexibility of models \cite{gordon2022jury, cabitza2023toward}. We provide a brief, non-exhaustive, review of prior perspectivist modeling approaches and applications.

\subsubsection{Direct Disagreement Prediction} One approach adopted in prior work is to directly model the variance in annotator labels as a function of features of the training instance (i.e., without using annotator features) \cite{raghu2019direct, gurari2017crowdverge}. \citet{gurari2017crowdverge} apply this approach to visual question answering tasks. They treat disagreement prediction as a binary task, using image- and question-based features to predict whether a panel of 10 raters will reach a supermajority (9/10 or 10/10) decision for a specific image-question pair. \citet{raghu2019direct} provide a theoretical analysis of direct disagreement prediction, arguing that disagreement prediction can be thought of as a regression task, mapping inputs to ``uncertainty scores,'' such as the variance of rater labels or the probability of two raters agreeing. They contrast direct disagreement prediction (referred to as direct uncertainty prediction in their work) with what they call uncertainty via classification: a two step-process in which an uncertainty score is computed from the output of a classifier that predicts the probability of a positive label. They prove direct disagreement prediction is more accurate than uncertainty via classification. 

\subsubsection{Annotator-Aware Approaches}  In contrast to direct disagreement prediction, annotator aware approaches utilize annotator-level features when making predictions \cite{gordon2022jury, fleisig2023majority, kumar2021designing, yin2023annobert, oluyemi2024corpus}. These features almost always include an annotator ID, for which corresponding embeddings are learned \cite{gordon2022jury, fleisig2023majority, yin2023annobert, oluyemi2024corpus}. Prior work differs on exactly how embeddings are learned, and where embeddings are incorporated in a neural architecture.
Demographic information for each rater is also sometimes utilized \cite{gordon2022jury, fleisig2023majority}, though ablation analysis from \cite{gordon2022jury} found limited additional value to using these features in a toxicity detection task. 

Practically speaking, an important distinction between direct disagreement prediction and annotator-aware approaches is the type of training data needed. Direct disagreement prediction requires multiple labels per training instance, but does not require any annotator identifiers features. In contrast, annotator aware approaches generally require annotator identifiers, but do not require multiple labels per training instance. 

\subsection{Towards Venire}\label{sec:background-summary}

The existing literature demonstrates that even well-intentioned moderation teams may suffer from undetected consistency issues \cite{lucas2019understanding, koshy2023measuring}, and that such consistency issues negatively affect user experience \cite{ma2022fairness, jhaver2019did}. Further, training a machine learning model to predict disagreements appears to be technically feasible \cite{gordon2022jury, gurari2017crowdverge}, making it possible to build the predictive model underlying Venire. However, we believe there are a few major open questions that need to be answered before building and testing Venire. First, how do moderators view the labor-consistency tradeoff inherent to panel review? Would they deem it worthwhile to increase the amount of decisions that need to be made in order to catch potential disagreements? And how could a panel review system complement existing moderation practices? Second, even if disagreement prediction is possible in some cases, is it possible to implement \textit{for a realistic content moderation task} and \textit{with the kinds of data available in a subreddit's moderation log}? These questions motivated our decision to conduct two preliminary studies leading up to building and evaluating the Venire interface. 
\section{Preliminary Interviews: Does Venire Support Moderator Needs?}\label{sec:prelim}

We conducted a round of preliminary interviews with moderators to better understand how they would view the labor demands of panel review. More broadly, we wanted to surface moderators' general attitudes towards the idea of an ML-assisted panel review system, and better understand moderators' existing processes for improving decision consistency. Simply put, our goal was to evaluate whether our vision for Venire aligned with moderator needs. We summarize these interview objectives in the following research questions:

\begin{itemize}
  \item[] \textbf{RQ1a}: How important do moderators think decision consistency is? What factors do they weigh it against?
  \item[] \textbf{RQ1b}: What processes do moderators currently employ to improve decision consistency?
  \item[] \textbf{RQ1c}: What benefits and harms do moderators anticipate an ML-guided panel review system will bring?
\end{itemize}

\subsection{Initial Prototypes}\label{sec:prelim-prototypes}

\begin{figure}
  \centering \fbox{\includegraphics[width=0.8\textwidth, keepaspectratio]{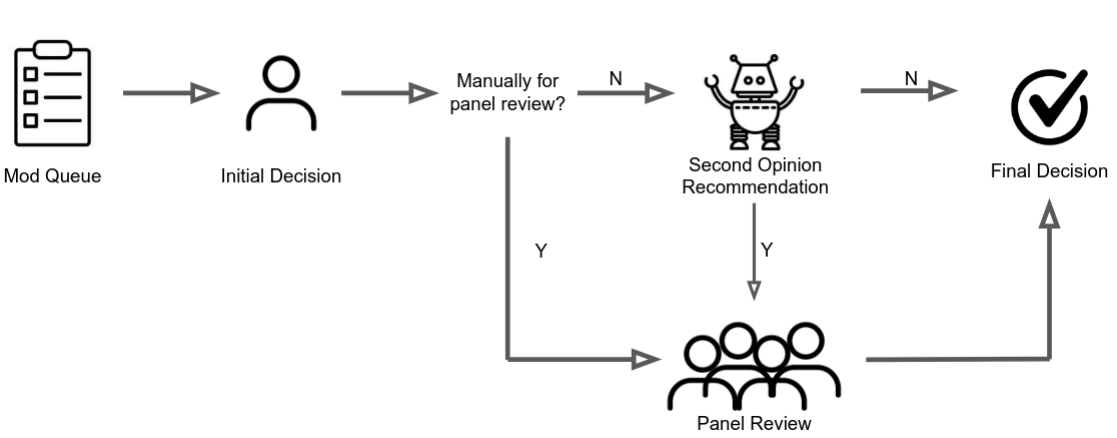}}
  \caption{Venire workflow. A case is pulled from the moderation queue by a human moderator, and an AI model recommends whether it should be reviewed by a single moderator or a panel of moderators.}\label{fig:workflow_diagram}
\end{figure}

\begin{figure}
  \centering \fbox{\includegraphics[width=\textwidth, keepaspectratio]{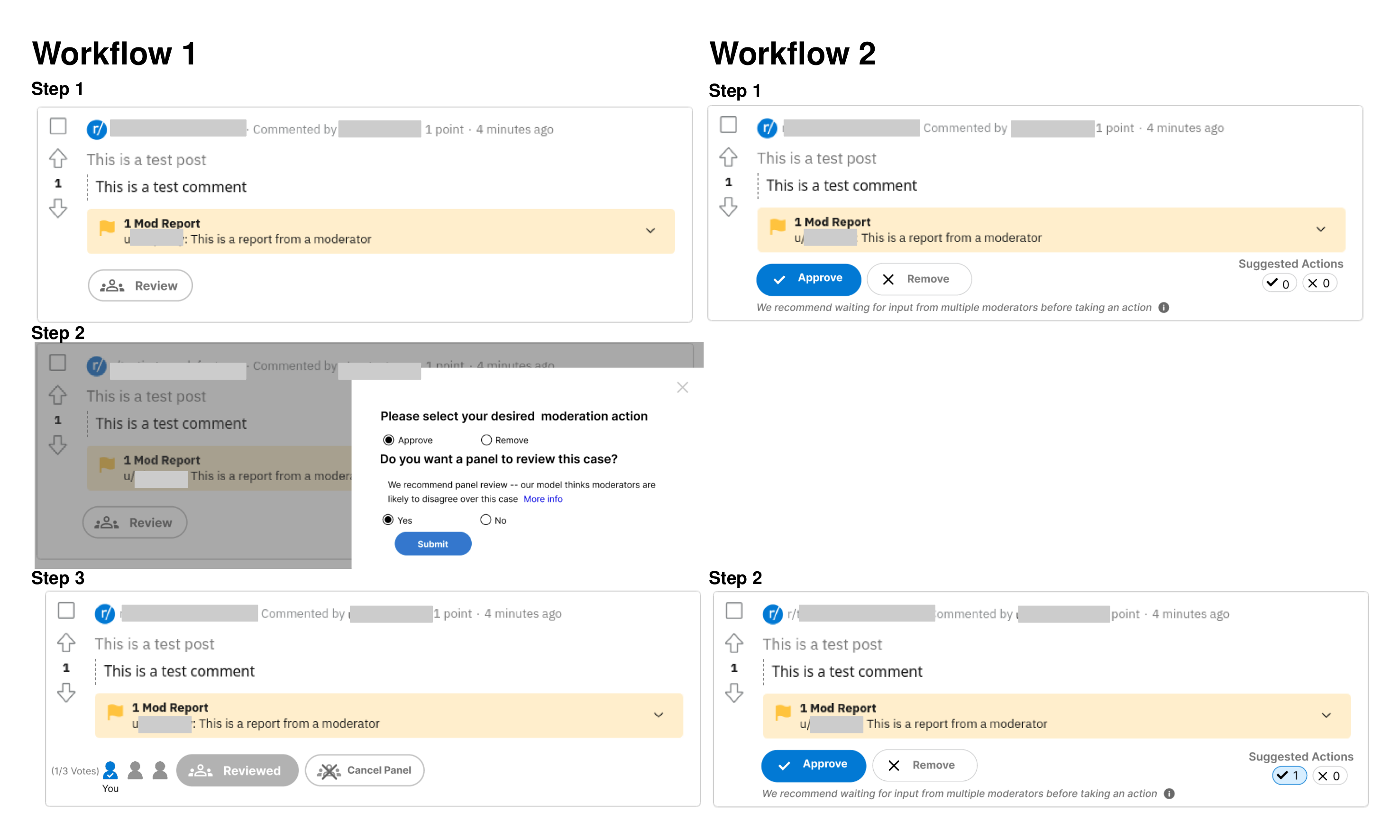}}
  \caption{Two potential Venire workflows. \textbf{Left: The strict voting workflow.} When making a ruling on a case, moderators must specify whether they want to flag it for panel review or not. Cases flagged for panel review remain in the moderation queue until a majority vote is achieved across $k$ raters. \textbf{Right: The suggested action workflow.} Rather than enforcing a strict voting procedure, moderators are always given the option to ``suggest'' an action instead of making a decision, making their opinion visible to other moderators. Any moderator can input a final decision when they feel confident enough. }\label{fig:mockups}
\end{figure}

To ground our interviews, we created a workflow diagram to communicate the idea behind our system (\Cref{fig:workflow_diagram}). We also created two interactive interface mock-ups in Figma (\Cref{fig:mockups}), representing more and less rigid versions of the panel voting process. In order to minimize disruptions to existing moderator workflows, we based the interface mock-ups heavily on the existing Reddit moderation queue interface. 

\begin{figure}
  \centering \includegraphics[width=\textwidth, keepaspectratio]{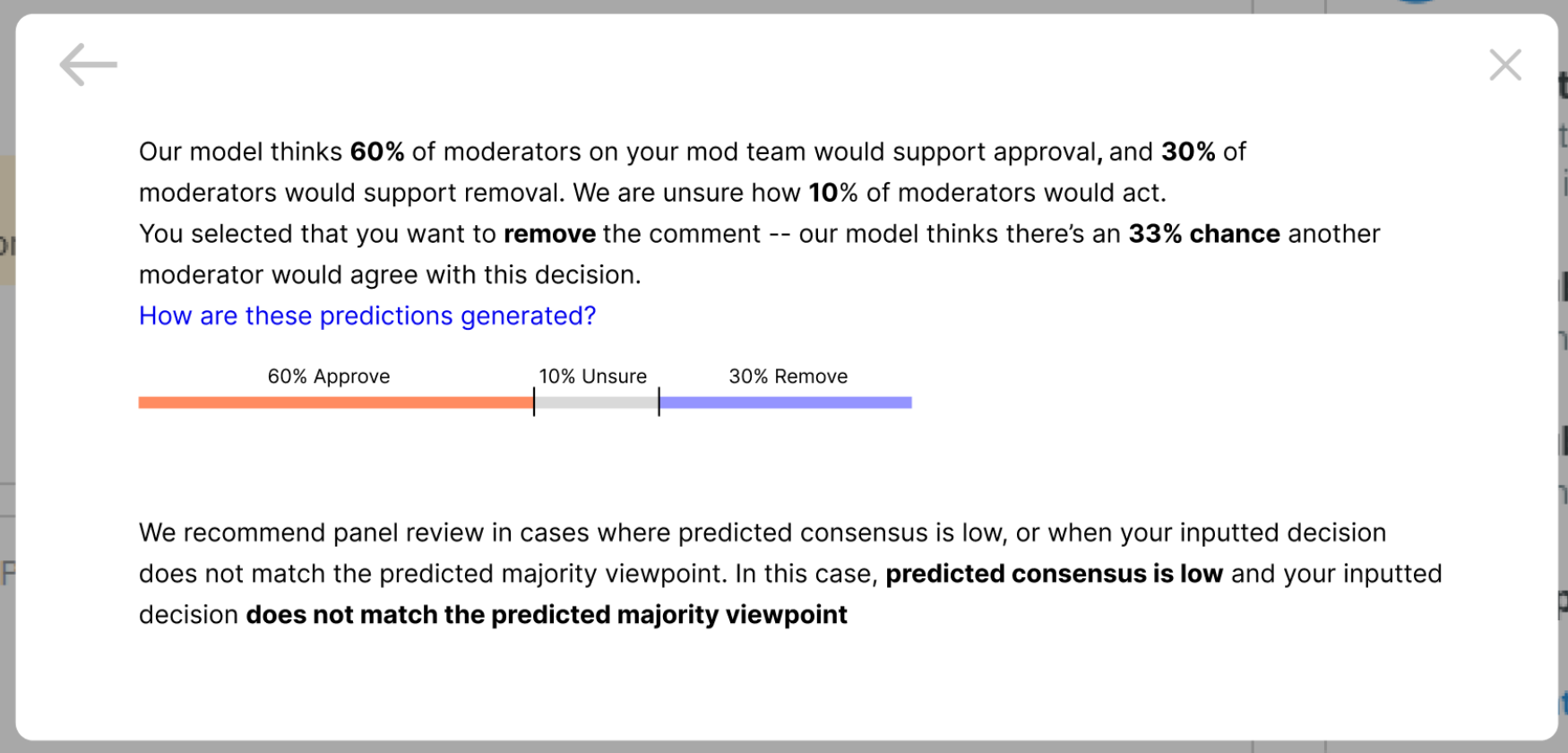}
  \caption{Visualization of how the model predicts the moderation team will react to a particular case---accessed by clicking the "More info" button next to the panel recommendation text in both mock-ups.}\label{fig:model-predictions}
\end{figure}


\subsubsection{System Mock-up \#1: Strict Voting}\label{sec:prelim-prototypes-1} \Cref{fig:mockups} contains the first version of the system we presented to moderators. In this version of the interface, moderators make two decisions for each comment: whether it should be removed or not, and whether it should undergo panel review or not. Moderators also see the model-generated recommendation for panel review based where appropriate. Clicking the "More info" button allows the user to see a full breakdown of how the model predicts the moderation team will react to the comment (\Cref{fig:model-predictions}). If the moderator chooses to flag a case for panel review, it will remain in the queue, until $k$ moderators cast a vote on it, where $k$ is a configurable number. After the $k$th vote is input, a decision will be determined based on the majority vote amongst moderators. Note that any moderators on the subreddit's mod team are allowed to weigh in on a panel decision---panel votes are not solicited from specific moderators. To minimize bias, moderators are unable to see the direction of existing votes until after they vote themselves.

\subsubsection{System Mock-up \#2: Suggested Actions}\label{sec:prelim-prototypes-2}  Rather than enforcing a majority vote across $k$ raters, the second interface treats panel review more loosely. All moderators are given the ability to signal their preference for removal or approva on the interface .There is no built-in aggregation mechanism for these suggestions. Instead, aat any time, a moderator can choose to make the final removal/approval decision, taking into account the other suggested actions as they see fit. When cases are predicted to be controversial, moderators are advised to make a suggestion rather than an immediate decision. This version of the system focuses on making the opinions of different moderators visible to one another, and aims to disrupt the existing moderation workflow as little as possible. 

\subsection{Interview Protocol}\label{sec:prelim-protocol}

Interviews were conducted over Zoom and lasted around 60 minutes each. Each interview consisted of three parts. First we asked moderators to describe the rules of the subreddit they moderate. We then asked moderators to recall any experiences they have had with disagreements over how to enforce the rules in their community, steps they have taken to ensure decision consistency (RQ1b), and general attitudes towards decision consistency as a goal for moderation (RQ1a). Next, we presented the workflow diagram to moderators to give them a high-level idea of how the system would work. Moderators were asked to speculate about the strengths and weaknesses of the system (RQ1c) and whether they thought it would be helpful in their community. If moderators did not explicitly mention concerns about increased workload here, we prompted them about it.

Finally, participants were given a guided tour of the two interactive mockups. Afterwards, we asked moderators to contrast the two interfaces, and state which one they preferred. We also solicited lower-level design feedback and asked moderators whether there were any additional benefits or harms from using the system they could imagine. Moderators were compensated the equivalent of \$30 USD in their local currency via Amazon giftcard or PayPal. 

\subsection{Recruiting}\label{sec:prelim-recruiting}

Recruitment messages were sent to Reddit moderation teams via the platform's modmail feature. Messages were sent to subreddits with over 10,000 subscribers and at least one subjective rule listed in the subreddit's sidebar. The lead author manually curated a list of subreddits based on these criteria. The r/ChangeMyView subreddit was also directly contacted since the researchers had access to a dataset of moderation decisions from this community that could later be used to train Venire's machine learning model. Several measures were taken to avoid spamming moderators with recruitment messages, which we detail in the \Cref{appendix:prelim-recruitment}. \Cref{tab:participant-info} contains a summary of the final eight participants interviewed.

\begin{table}[ht]
\centering
\begin{tabular}{|c|c|c|}
\hline
\textbf{PID} & \textbf{Subreddit Size (Subscribers)} & \textbf{Subreddit Topic} \\ \hline
P1          & 1-10M                                & r/ChangeMyView           \\ \hline
P2          & 10-100K                              & Fanfiction               \\ \hline
P3          & 100K-1M                              & Gaming/TTRPG             \\ \hline
P4          & 10-100K                              & Music                    \\ \hline
P5          & 10-100K                              & Pregnancy Support        \\ \hline
P6          & 100K-1M                              & Gardening/Agriculture    \\ \hline
P7          & 100K-1M                              & History/Politics         \\ \hline
P8          & 10-100K                              & TV Show                  \\ \hline
\end{tabular}
\caption{Interview Participant Information}\label{tab:participant-info}
\end{table}

\subsection{Results}\label{sec:prelim-results}

\subsubsection{Attitudes Towards Decision Consistency (RQ1a)}\label{sec:prelim-results-rq1a}

Unsurprisingly, almost all the moderators we interviewed believed moderation decision consistency to be desirable (N=6). P3, speaking directly to the motivations of the project, stated ``in an ideal world [...] every post would go through panel review.'' This was especially the case for subreddits that dealt with highly sensitive or political topics (N=3). P1, who moderates r/ChangeMyView, argued that decision consistency was essential to user participation in their community:
\begin{quote}
    ``Ultimately the subreddit that we have doesn't work unless there is consistent moderation that is as topic neutral as possible. [...] Pick any culture wars topic in the United States or any hot button political issue---keeping conversations civil on those topics requires a lot of consistency. Everyone needs to know what the rules are and how they're applied. And they need to feel it's fair and consistent regardless of if they're a Republican, a Communist, a Social Democrat, a Centrist, or a Libertarian. They need to feel like they're getting a fair shake or they don't participate.''
\end{quote}
Still, most moderators contextualized the importance of decision consistency alongside other factors (N=5). Minimizing workload and stress came up most frequently (N=4). With regards to stress, P5 said that moderation could sometimes feel like ``factory work'', and that it was important to "get through the queue'' with the ``least amount of damage to yourself.'' Other moderators highlighted the voluntary nature of moderation work when thinking about the consistency-workload tradeoff. P2 argued that ``whoever's the person shouldering most of the workload gets to make the calls'' in part because those moderators would be ``more in tune with how the subreddit currently is.'' Similarly, P4 said ``we really try to back up our moderators as much possible in their decisions, even if it's something we might disagree with.''

Outside of workload concerns, P3 mentioned that incorporating diverse perspectives into the moderation team could be worth sacrificing some decision consistency for. Additionally, P2 noted that more senior moderators were sometimes able to take actions within the community that other moderators were not able to. This is a form of decision inconsistency that was viewed positively. They described these moderators as ``having a bit more goodwill'' amongst community members, which allowed them to ``shut something down'' where another moderator might not have been able to. 

\subsubsection{Existing Practices (RQ1b)}\label{sec:prelim-results-rq1b}

Every moderator we interviewed described taking steps in the past to either preempt a potential disagreement (N=8) or resolve a disagreement that surfaced after a moderation action had already been taken (N=5). Most moderators recalled soliciting a second opinion from another moderator through a side channel like Discord (N=5). This practice was sometimes specifically encouraged for new moderators (N=2). Moderators also described holding discussions prior to implementing a new rule to try to iron out potential inconsistencies (N=3). In a few cases, moderators outlined specific rules or policies where taking a vote amongst multiple mods was required (N=2). In P1's community, certain kinds of post removals  ``required two moderators to sign off on.''

Still, most moderators felt disagreements were relatively rare to begin with (N=5). At first glance, this was surprising given the disagreement rates found in prior work. However, the moderators we interviewed attributed the rarity of disagreements to the fact that their subreddits only had a few moderators (N=2) or to the fact that their community's rules were straightforward (N=2). In contrast, the subreddits studied in prior work had large moderation teams, and tight moderation standards. Still, P1 mentioned that disagreements could be going unnoticed in their community, stating: ``if a moderator does make a decision, very rarely are the other moderators going to even be aware of it.'' 

\subsubsection{Potential Benefits and Harms of ML-Guided Panel Review (RQ1c)}\label{sec:prelim-results-rq1c}

Moderators outlined a number of potential benefits to a panel review system. Even without the predictive model, moderators appreciated having a built-in channel for disagreement-resolution that might have otherwise happened in a side channel (N=5). And as we anticipated, a number of moderators felt that a machine learning model could help surface disagreements that would have gone unnoticed (N=6). P1, for example, said ``I would say this tool would be great for helping to figure out [...] if there are controversial cases that are being decided too quickly.'' A few moderators explicitly stated that catching these disagreements could lead to policy updates (N=3). P7 was one such moderator:

\begin{quote}
    ``You might think that you're in the majority with one opinion or someone else might think they're in the majority with another opinion and they're not. So it would be nice to be able to see, 'oh, this is how the rest of the mod team has been moderating things. I see that I've been moderating differently.' Maybe we should talk about this rule and it will provide discussion and at least get all the mods on the same page or maybe a compromise to the rule is made.''
\end{quote}

 We also surfaced a few unanticipated benefits. Most notably, moderators highlighted the benefits of ML-guided panel review as an onboarding tool (N=4). In the context of recruitment, P2 argued that this would be beneficial to both senior and new moderators. For the senior moderators, P2 felt that they ``won't feel like they have to be double checking or [...] correcting them all the time.'' and for the moderators being onboarded ``they can be a little bit more confident that if they take an action and it turns out to be a mistake, that it will be caught. Like a safety net.'' P5 echoed this sentiment saying ``I do think that you'll have people more willing to moderate in general. So right now when you recruit moderators, the stickiness of a moderator is not high. [...] It could be that when they run across these difficult to moderate content, they don't know what to do.'' P3 contrasted using the panel review system to their current practices for onboarding new moderators, saying:

 \begin{quote}
     ``When there's a new moderator on the team, I'll try and do audits of some of their actions. But it's a bit of a slog and there aren't really great tools to do that. Whereas this serves as a potential to be doing those audits in a way. And so that just makes it easier for all of us''
 \end{quote}

However, moderators mentioned a number of drawbacks that are important to consider. Workload concerns came up multiple times (N=5), especially with respect to false positive flags from the ML model (N=3). At the same time, moderators felt like the workload could be manageable if disagreements were relatively rare (N=4). One participant, P5, questioned the fundamental value of highly deliberative moderation at all, saying:

\begin{quote}
      ``I don't always 100\% agree with everything the other mod does but I agree enough, and that's enough for me. And that's because moderating a community is a lot of work. I have a full-time job, I have a kid. There's other things happening in my life. And in hindsight over the six years, I think the community is better for me not strictly creating my own vision [..] Building consensus is a difficult thing to do, and it's not always worth the bang for the buck. It's not always worth it for you internally as a person, and it also might not be worth it for the community in general.''
\end{quote}

Moderators also felt that panel review could increase the time it takes for a final decision to be reached on a post (N=3). P6, for example, said ``if it takes three days to get an answer [...] the post is gone [...] it doesn't really even matter anymore.'' However, P2 offered a potential solution saying the system could instead be ``more of an appeals process that would let you reverse a decision [...] rather than something that might block you from taking action.''

Finally, a few moderators speculated that the panel review feature might simply go unused (N=4). P3 and P6 both argued that moderators may not have the self-awareness to flag a case for panel review. P6 stated ``it takes a certain level of person to say 'I think it's possible that I didn't make the best decision here.' '' P3 argued one solution might be to have the ML model force panel review rather than merely suggesting it. Other moderators simply felt that moderators within the same mod team might all agree anyways (N=2). To mitigate this, P6 proposed that instead the system could get ``an outside panel of moderators'' instead, possibly from a ``sister community.'' 

\subsubsection{Mockup Preference}\label{sec:prelim-results-preference}

Moderators largely preferred the first strict voting mock-up to the suggested action mockup (N=5 vs N=1). Moderators who preferred the strict voting mock-up liked the rigidity of the voting process. These moderators described the strict voting mock-up as ``more formal'' and the suggestion action mock-up as ``more passive'' and ``softer.'' P2, the sole moderator who preferred the suggested action mockup argued that they liked the ``non-binding'' nature of the suggested action button. They felt the voting system ``puts a bit more pressure on you'' because ``if you're the second person voting, you might or might not be making a moderator action, you don't know.''

\subsubsection{Summary of Findings} 

Overall, we find that Venire's intended goals, surfacing disagreements and improving decision consistency, are aligned with moderators' values. Still, moderators reported a number of additional factors, like stress management and decision speed that these goals should be weighed against. Although moderators reported employing practices to improve decision consistency already, they saw potential for Venire to further these efforts. Thus, our findings gave us the confidence to proceed with building Venire. 

\section{Technical Evaluation: Can Venire Improve Decision Consistency?}\label{sec:technical}

In this section, we present the results of two experiments that test the technical feasibility of predicting moderation disagreements. Our primary goal is to demonstrate that ML-guided panel review can deliver on the promise of improving decision consistency and surfacing disagreements. In our first experiment, we demonstrate this using a large, publicly available toxicity dataset. In our second experiment, we construct a new, more ecologically valid dataset using crowdworkers on Prolific, and repeat our analysis. Crucially, the training dataset used in the second experiment is much smaller, and based on a more realistic content moderation task. We find that our model performs worse on this new dataset, but still allocates panels more effectively than a random panel assignment baseline. More concretely, the goal of our technical evaluations is to answer the following research questions: 

\begin{itemize}
    \item[] \textbf{RQ2a}: To what extent can an ML-guided panel review improve the consistency of moderation decisions?
    \item[] \textbf{RQ2b}: To what extent can an ML-guided panel review surface disagreements between moderators?
    \item[] \textbf{RQ2c}: How does the performance of an ML-guided panel review system change when we move from an ideal dataset to one that more closely matches moderation log data?
\end{itemize}

\subsection{Modeling Approach}\label{sec:technical-model}

\begin{figure}\label{fig:architecture}
  \centering \includegraphics[width=0.8\textwidth, keepaspectratio]{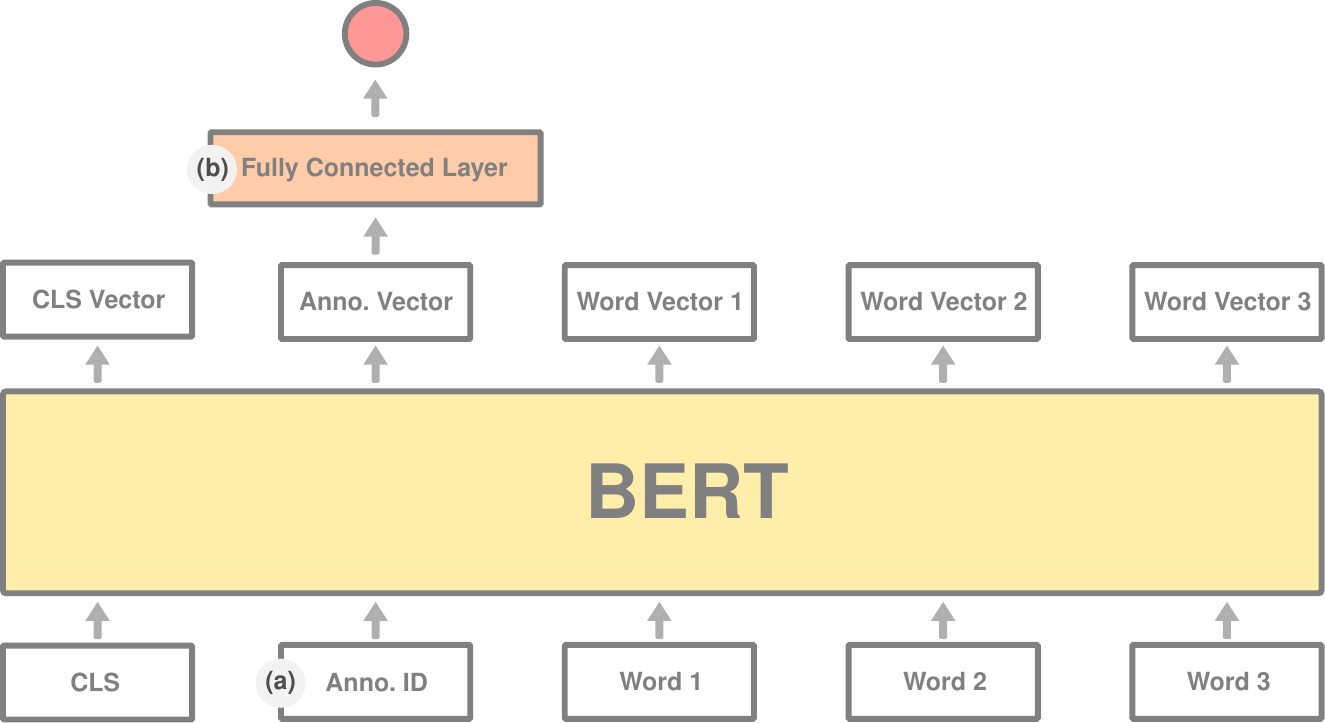}
  \caption{Overview of our model architecture. We prepend a unique token representing a specific annotator (a) to each sentence, and pass the finetuned BERT contextual embedding for this token through a fully connected layer (b) to produce our final prediction. We ignore the special CLS token that is typically used for BERT sequence classification tasks \cite{devlin2018bert}.}
\end{figure}

Our modeling approach is inspired by recent perspectivist NLP papers \cite{gordon2022jury, fleisig2023majority, yin2023annobert, oluyemi2024corpus}. We define the prediction task as follows. As input, the model is shown a text $x_i$ and the identity of a rater $j$. Our model then outputs a prediction for how rater $j$ would label that $x_i$. Prior work varies on the specific neural architecture used. In preliminary experiments, we found adapting Yin et al.'s approach to be most effective \cite{yin2023annobert}. We treat each rater ID, $j$, as a special token that is prepended to $x_i$. We feed this new string into a BERT model, and pass the contextual embedding for the rater ID token through a feedforward layer to produce the final prediction. \Cref{fig:architecture} demonstrates our proposed architecture. Like other perspectivist approaches \cite{gordon2022jury, fleisig2023majority} we learn rater-specific embeddings to capture attributes of decision-making style. This embedding is passed through multiple transformer layers alongside the text, to produce a final contextual embedding which captures attributes of the text and the rater.

During deployment, our model makes a prediction about how \textit{each} possible rater would label a $x_i$. These predictions can be aggregated to produce: 1) a ``majority vote'' prediction, $M(x_i)$, of the fraction of raters that would label $x_i$ as positive, and 2) a ``disagreement score'' $D(x_i)$, that captures how controversial $x_i$ is. To produce $D(x_i)$, we calculate the empirical probability that two randomly selected rater predictions disagree with one another \cite{raghu2019direct}. It is natural to ask why we go through the process of producing a rater-specific prediction in the first place---an alternative approach might be to simply train a model that learns $D(x_i)$ directly. In the context of our study, this alternative approach is infeasible, since it requires us to have multiple rater labels for each comment in the training dataset; real moderation log data is largely singly-labeled.   

\subsubsection{Hyperparameters} We use a base BERT model for the toxicity prediction task and a BERT-large model for the Prolific study. Because the comments for the toxicity prediction task are short, we use only the first 126 tokens of the text for prediction. For the Prolific study, we use a 256-token window. When comments exceed the token window, we create an embedding for each 256-token slice of the comment and combine them via max-pooling to produce a final token embedding. Additionally, for the Prolific task, we also feed the model the first 256 tokens of text of the immediate parent comment for reply comments, and the first 256 tokens of the post body for top-level comments. Unfortunately, this additional thread context was not available in the toxicity dataset. We separate the target comment's text from the thread context using BERT's SEP token. Our model is implemented in PyTorch and trained with the AdamW optimizer---we use a learning rate of 2e-5, batch size of 32, and weight decay of 0.0075. We train for 3 epochs on the toxicity dataset and 5 epochs on the Prolific dataset. The hyperparameters were determined after conducting a modest grid search using a validation set. All model training was conducted on a single Nvidia T4 GPU provided by Google Colab. 

\subsection{Experiment 1: Toxicity Dataset}\label{sec:technical-toxicity}

\subsubsection{Dataset Description} For our initial evaluations, we leverage a large, multiply-labeled toxicity dataset provided by \citet{kumar2021designing}. This dataset contains 107620 comments sampled from Twitter, Reddit and 4chan. Each comment received five, 5-point Likert ratings of toxicity. A total of 17280 raters contributed to the dataset, and each rater labeled at least 20 comments. For our evaluation, we treat toxicity prediction as a binary classification problem, treating Likert ratings of 3 ("moderately toxic") or higher as "Toxic", and 2 ("slightly toxic") or lower as "Not toxic" \cite{kumar2021designing}. The labels in this dataset are mildly imbalanced---only 29\% of supplied ratings were "Toxic", and the majority vote was "Toxic" for only 21\% of comments.  We use 75\% of the comments in the dataset for training, 10\% for validation, and reserve 15\% for reporting results.

\begin{table}[ht]
\centering
\begin{tabular}{|l|c|c|c|c|}
\hline
\textbf{Model} & \textbf{AUROC} & \textbf{Accuracy} & \textbf{Precision} & \textbf{Recall} \\ \hline
\textbf{Toxicity (Rater-level Annotations)} & 0.8798 & 0.8191 & 0.7102 & 0.6445 \\ \hline
\textbf{Toxicity (Majority Vote)}           & 0.9196 & 0.8688 & 0.7045 & 0.6780 \\ \hline
\textbf{Disagreement (Rater-Aware)}         & 0.7376 & 0.7338 & 0.6373 & 0.4480 \\ \hline
\textbf{Disagreement (Rater-Blind)}         & 0.6946 & 0.7004 & 0.5453 & 0.4261 \\ \hline
\end{tabular}
\caption{Model prediction quality for the toxicity dataset}\label{tab:toxicity-acc}
\end{table}

\subsubsection{Toxicity Predictions} The first two rows of \Cref{tab:toxicity-acc} demonstrate our model's ability to predict toxicity labels. Our model achieves an accuracy of 82\% (AUROC 0.88) when predicting individual annotator ratings (using a threshold of 0.5), and 86\% (AUROC 0.92) when predicting the majority vote amongst all five annotators. Majority vote predictions are made by producing a binary prediction for each of the five annotators and taking a majority vote amongst predictions.

\subsubsection{Disagreement Predictions} To test our model's ability to predict disagreements amongst raters, we divide comments in the test set into two groups---high consensus comments (decided unanimously or with a single dissenting rater) and low consensus comments (decided by a 3/2 split amongst the raters). Under this definition, 30\% of comments were considered low consensus. The goal of the disagreement prediction task is to discriminate between these two classes. We produce two kinds of predictions for comment consensus-level: "annotator blind" predictions and "annotator aware" predictions. As the name suggests, annotator-blind predictions are produced without looking at the identities of the five raters who were assigned to the comment. Instead, we randomly sample 100 annotators from the training set, and produce a prediction for each one. We then aggregate these 100 predictions into a single disagreement score (as described in \Cref{sec:technical-model}), and threshold the disagreement score to produce a final binary consensus-level prediction. This threshold is chosen by calculating the disagreement score at which a low consensus outcome is more likely than a high consensus outcome, assuming the binary annotator-level predictions are correct. To produce annotator-aware predictions, we perform the same procedure, but use the 5 raters actually assigned to the comment rather than a random sample. The second two rows of \Cref{tab:toxicity-acc} contain the results---annotator-blind predictions are 70\% accurate (0.69 AUROC), while annotator-aware predictions are 73\% accurate (0.74 AUROC).

\subsubsection{Simulation Analysis} 

While the previous results give us insight into our model's raw predictive power, they do not tell us how well our model will perform \textit{when used to allocate human decision-makers}. To address this, we present the results of a simulation based analysis using the test set data. First, we simulate single moderator review by randomly selecting a label, $h_{initial}(x_i)$ from one of the $n$ assigned raters for each comment $x_i$ in the test set. After simulating initial decisions, we then apply a \textit{panel allocation strategy} to the comments. Intuitively, a panel allocation strategy is a function that looks at $x_{i}$ and $h_{initial}(x_i)$, and triages the case for panel review. More formally, a panel allocation strategy is a function that takes in as input: $x_i$, $h_{initial}(x_i)$, and a list of model-generated predictions $f_j(x)$ for how each human rater $j$ will respond to $x_i$. For each case, a panel allocation strategy outputs panel priority score $p_i$. We then assign the cases with the top k\% highest $p_i$ to panel review. For cases assigned to by panel review we solicit a second, randomly selected opinion $h_{second}(x_i)$. If $h_{second}(x_i)$ does not match $h_{initial}(x_i)$ we solict a third vote, $h_{third}(x_i)$ to tie-break. 

We judge the performance of a panel allocation strategy by looking at 1) the amount of labor it required (average number of raters used per case), 2) the resulting decision consistency (how often the final decision matched the ground truth majority vote amongst all five raters), and 3) the number of disagreements it surfaced (how often did the panel contain at least one dissenting opinion). It should be noted that 2) and 3) represent distinct goals for the system. A panel allocation strategy that deliberately oversamples minority opinions, for example, might surface a large number of disagreements, but fail to improve decision-consistency. In a real world deployment, surfacing disagreements and improving consistency are more closely related, since moderators who observe disagreements can update community policy to minimize future disagreements. We cannot test with our simulation analysis, so we report disagreements surfaced and decision-consistency as separate measures. We test the following strategies for panel assignment:

\begin{itemize}
    \item[] \textbf{Random Panel Assignment}: The panel priority score is a random number between 0 and 1. 
    \item[] \textbf{Predicted Majority-Based Assignment}: The panel priority score is equal to $| h_{initial}(x_i) - M(x_i)|$. Intuitively, cases are prioritized for panel review when the initial human rater decision disagrees with the predicted consensus viewpoint. We use a random sample of 100 training dataset raters to produce $M(x_i)$. 
    \item[] \textbf{Predicted Disagreement Score-Based Assignment}: The panel priority score is equal to  $D(x_i)$, the probability that the predicted decisions for two out of 100 randomly sampled training dataset raters disagree. 
    \item[] \textbf{Predicted Disagreement+Majority Based Assignment}: The panel priority score is equal to the $2D(x_i) + M(x_i)$. We multiply $D(x_i)$ by two since $0 \leq D(x_i) \leq 0.5$, while $0 \leq M(x_i) \leq 1$
\end{itemize}

\begin{figure}
  \centering \includegraphics[width=\textwidth, keepaspectratio]{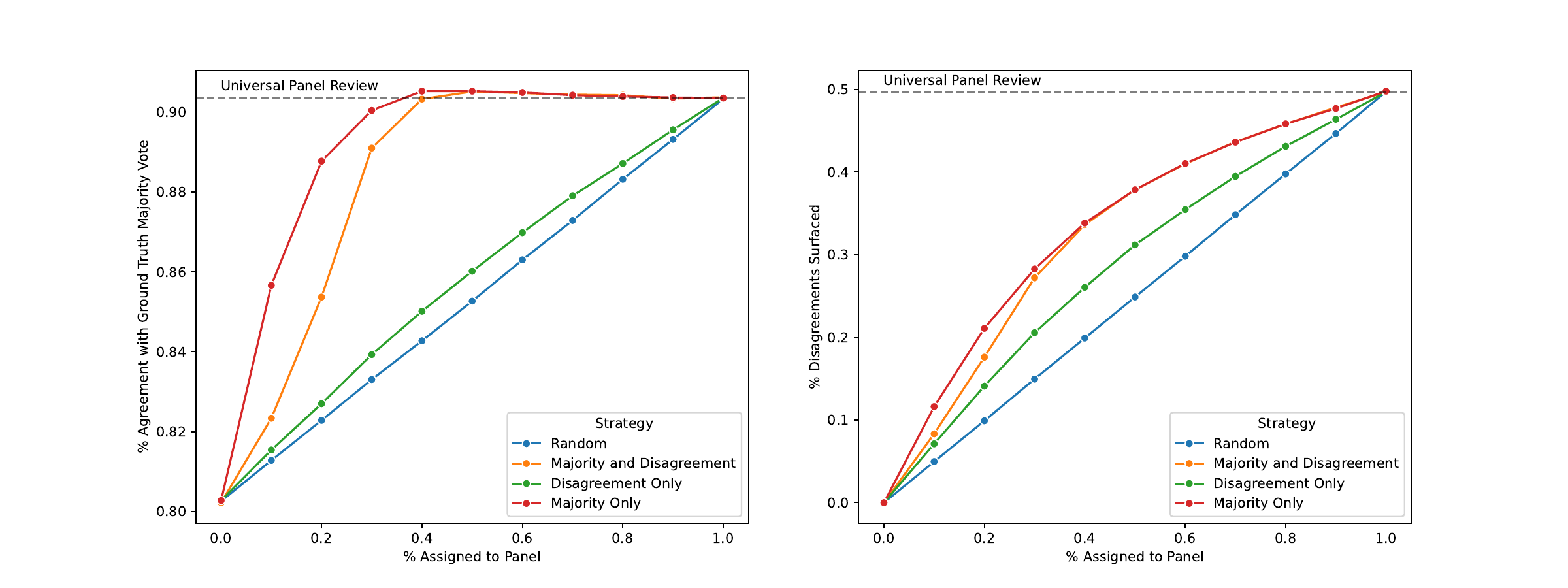}
  \caption{Performance of a panel prioritization strategies at increasing workloads. When it comes to improving decision consistency (RQ2a), majority-vote based prioritization converges to the optimal value much faster than random allocation. Majority-vote based prioritization also outperforms other strategies at surfacing disagreements(RQ2b), though to a less dramatic degree}\label{fig:toxicity-simulation}
\end{figure}
    
\Cref{fig:toxicity-simulation} contains the results of applying each strategy to increasing proportions of the test set. In each plot, the dotted line indicates the performance of assigning every case to a three person panel. Intuitively, effective strategies will quickly converge to the dotted line---these strategies approximate the benefits of universal panel review with fewer extra ratings. Overall we can see that Predicted Majority-Based Assignment performs the best---this strategy approximates the decision consistency benefits of full panel review after just over 30\% of cases are assigned to panel. This finding matches our initial intuition: many cases are straightforward, and do not benefit from undergoing panel review. Predicted Majority-Based assignment effectively identifies the subset of cases where panel review is most beneficial. All strategies outperform Random Panel Assignment for both surfacing disagreements and improving decision consistency. Somewhat counter-intuitively, the Predicted Disagreement Score-Based Assignment does not seem to improve decision consistency, contradicting suggestions from prior work \cite{raghu2019direct}. This makes more sense on closer inspection---when soliciting additional opinions on a highly controversial case, we are almost equally likely to sample the minority and majority viewpoint. Starting a panel only marginally improves the chance that we arrive at the ground truth majority viewpoint. Still, disagreement-based panel allocations outperform random assignment when it comes to surfacing disagreements. 

\subsection{Experiment 2: Prolific Dataset}\label{sec:technical-prolific-protocol}

Our initial analysis provides evidence that an ML-guided panel review system can improve decision consistency (RQ2a) and surface disagreements (RQ2b). However, the toxicity dataset we used differed from real moderation log data along a couple key dimensions, prompting us to experiment further (RQ3c). Perhaps most importantly, the dataset contained labels from multiple raters for each comment in the training dataset. Data scraped from a subreddit's moderation log will largely contain only a single decision per comment, since moderators do not usually re-evaluate old moderation cases. Intuitively, this could make the disagreement prediction task easier, since the model is able to learn explicit contrasts between decision makers during training. Additionally, the raw number of comments in the training dataset was high---only the largest communities on Reddit would have such large-scale log data. Finally, the toxicity labeling task is too broad to serve as an effective proxy for a content moderation task. The survey instrument used to produce the dataset did not clearly define toxicity \cite{kumar2021designing}. This may inflate the number of disagreements in the dataset when compared to a more clearly defined community rule. 

To address these issues, we curated a smaller dataset using crowdworkers on Prolific. To make the moderation task more realistic, we chose to have crowdworkers apply a real community rule to comments from a subreddit's actual moderation log. Specifically, we chose the r/ChangeMyView subreddit's "Rule 2", which bans comments that express rudeness or hostility towards another user. This rule was chosen for a few reasons. r/ChangeMyView maintains detailed guidelines on how Rule 2 should be applied. This makes it easier for us to communicate the rule to crowdworkers in an unambiguous manner, while ensuring ecological validity of the content moderation task. At the same time, Rule 2 still contains a degree of subjectivity. The Rule 2 guidelines outline a number of "edge cases" that require moderators to make a judgment call, like determining whether a comment contains "excessive passive-aggression."  Thus, it is reasonable to assume that there are disagreements that can be surfaced in the first place. Finally, Rule 2 requires relatively little thread context to moderate. This allows us to keep the cognitive load of the task low, since users will not read long threads of discussion in order to make a determination. 

In our Prolific study, crowdworkers contributed model training and test set labels across four separate tasks. Full details about the survey instruments are provided in the \Cref{appendix:prolific-materials}, but we provide an overview here. In the first task, participants were given a short introduction to the r/ChangeMyView subreddit and shown a set of condensed guidelines for how to apply Rule 2. Participants then completed three practice questions where they were asked to apply Rule 2 to a comment. In these practice questions, participants were shown the actual decision made by moderators alongside a short explanation. The practice questions were deliberately chosen to contain short, straightforward cases that illustrated key points in the Rule 2 guidelines. Participants were then asked to provide training set labels for 20 comments and test set labels for 20 comments. Participants were compensated \$7.50 for completing the first task (20 minute average completion time). The subsequent tasks each began with a refresher on the Rule 2 guidelines. In tasks 2 and 3, participants provided 20 training set labels and 20 test set labels. In task 4, participants provided 40 training set labels. Participants were compensated \$5 for each of these parts (average completion times of 18 minutes for parts 2 and 3, and 20 minutes for part 4). This structure was chosen to ensure that we would still have sufficient training and test data for each participant, even if they dropped out before completing all four parts. To minimize attrition between tasks, participants received a \$22.50 bonus for completing all four tasks, effectively doubling their compensation. 

Participants were screened to include only participants who were: US Residents, first-language English speakers, between the ages of 18 and 65, and Reddit users. In total 34 participants completed at least one of the tasks, and 32 participants completed all four tasks. Below, we provide a few additional details about how training and test labels were solicited.

\subsubsection{Training Labels}

\begin{figure}
  \centering \fbox{\includegraphics[width=0.5\textwidth, keepaspectratio]{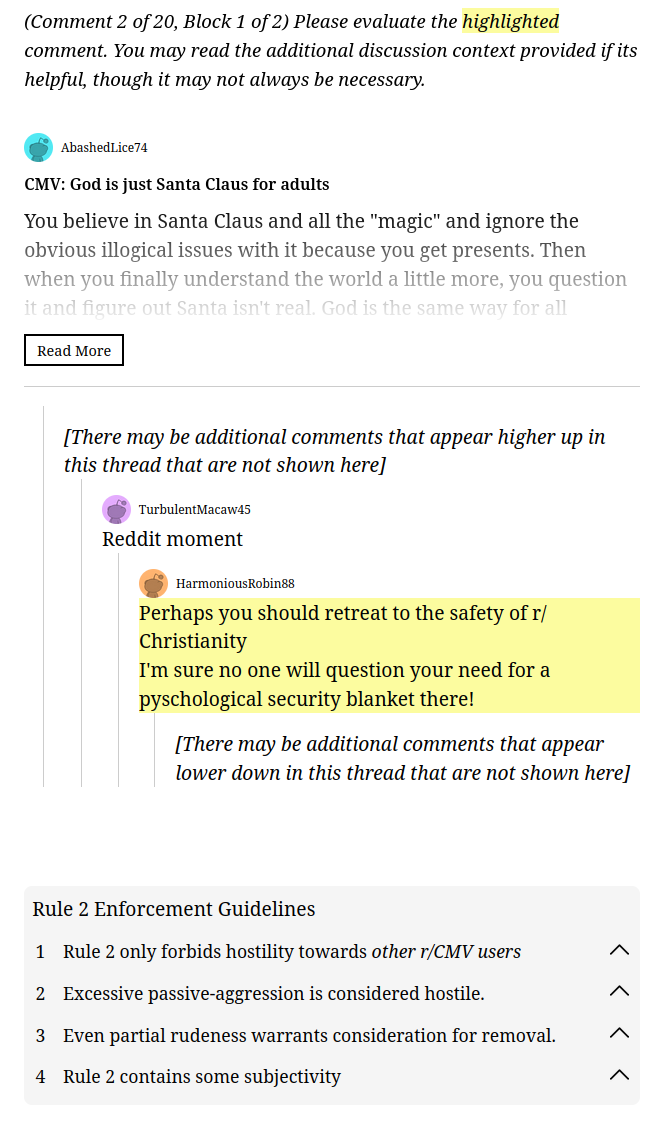}}
  \caption{Survey interface recreation of a comment. In addition to providing the text of a comment, we also include a parent comment (where appropriate), and the post associated with the comment.}\label{fig:comment-recreation}
\end{figure}

\Cref{fig:comment-recreation} demonstrates our comment interface. We display a comment's text, the text of any immediate parent comment, and the title and body of the associated post. Usernames are replaced with pseudonyms. Our interface also contains a collapsible version of the Rule 2 guidelines. Participants were given the following prompt to solicit labels: "Please state whether you believe Rule 2 (Banning rudeness/hostility towards other users) applies to the highlighted comment." The answer choices were: "The comment violates Rule 2"/"The comment does not violate Rule 2." To mimic moderation log data, each comment in the training dataset was shown to a single survey-taker. Participants who completed all four tasks provided 100 training labels. A few participants experienced technical issues with the survey and supplied a different number of labels (ranging from 80 to 160). In total 3280 comments were labeled for the training dataset.

\subsubsection{Test Labels}

Each comment in the test dataset was shown to multiple participants. We used the same comment recreation interface for test set comments. We decided to use the test set as an opportunity to assess how well human raters were able to anticipate disagreements, creating a baseline for model performance. To do this, we informed participants that comments in the test set section of the survey were being shown to 6 other human raters. Then, in addition to asking them to apply Rule 2, we asked participants to predict whether the other 6 raters would be in "high consensus" or "low consensus" about whether Rule 2 applied. High consensus cases were those where a super-majority of 5 or more raters ruled in favor of one side. To incentivize accurate predictions, participants received a small bonus of 10 cents for each correct prediction (maximum of \$2 bonus for a single task). Under this scheme, we need 7 human-rater labels per comment. Due to a technical issue with our sampler, we ended up soliciting 10 labels per comment instead, giving us a few extra. In total 200 comments were labeled for the test dataset. 

\subsection{Prolific Study Results}\label{sec:technical-prolific-results}

\begin{table}[ht]
\centering
\begin{tabular}{|l|c|c|c|c|}
\hline
\textbf{Model} & \textbf{AUROC} & \textbf{Accuracy} & \textbf{Precision} & \textbf{Recall} \\ \hline
\textbf{Rule 2 Application (Rater-level Annotations)} & 0.8163 &  0.7345 & 0.7515 & 0.7933 \\ \hline
\textbf{Rule 2 Application (Majority Vote)}           & 0.8540 & 0.7692 & 0.7425 & 0.8794 \\ \hline
\textbf{Disagreement (Rater-Aware)}         & 0.6197 & 0.6028 & 0.4909 & 0.4204 \\ \hline
\textbf{Disagreement (Rater-Blind)}         & 0.6295 & 0.6154 & 0.5101 & 0.4560 \\ \hline
\textbf{Disagreement (Human-Generated Predictions)}         & N/A & 0.6212 & 0.5219 &  0.4005 \\ \hline
\end{tabular}
\caption{Model prediction quality for the Prolific dataset}\label{tab:prolific-acc}
\end{table}

\subsubsection{Rule 2 Application Predictions} As in \Cref{sec:technical-toxicity}, we present our model's ability to predict individual annotator labels and majority vote decisions. The first two rows of \Cref{tab:prolific-acc} demonstrate our model's ability to predict individual rater Rule 2 determinations. Our model achieves an accuracy of 73\% (AUROC 0.82) when predicting individual annotator ratings (using a threshold of 0.5), and 77\% (AUROC 0.85) when predicting the majority vote amongst all annotators. Unsurprisingly given the size of the respective training datasets, our model performs slightly worse on this task compared to the toxicity task.

\subsubsection{Disagreement Predictions} Cref{tab:prolific-acc} also contains the results of the disagreement prediction tasks. We also include the quality of the human supplied disagreement predictions. When computing disagreement prediction performance, we randomly select a rater for each comment, and compare their disagreement prediction (and the model's prediction) against the Rule 2 determinations of a random sample of 6 other raters. We report the average of 100 simulations. Again, we see that the performance of the model is generally worse when compared against the toxicity dataset (62\% accuracy vs 73\%). Still, the rater-blind model is able to make disagreement predictions at roughly the accuracy of human raters (62\% for both). 

\subsubsection{Simulation Analysis} 

\begin{figure}\label{fig:prolific-simulation}
  \centering \includegraphics[width=\textwidth, keepaspectratio]{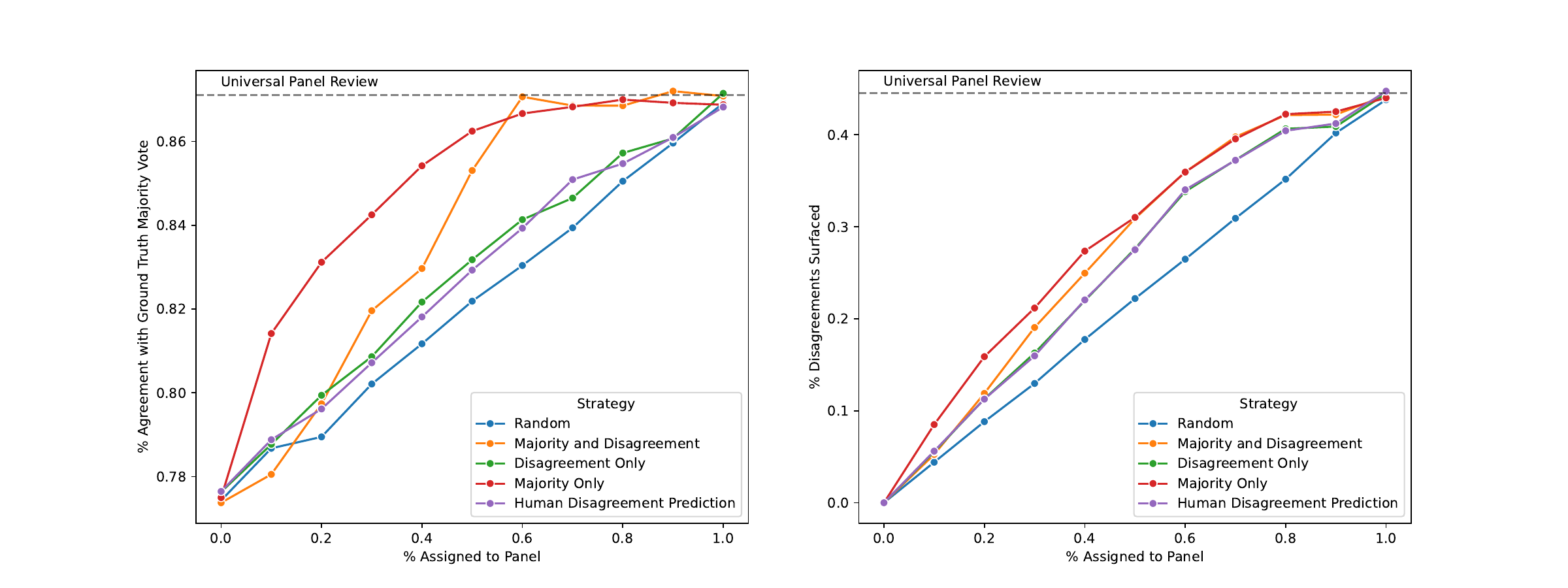}
  \caption{Performance of a number of panel prioritization strategies as increasing workloads. When it comes to improving decision consistency, majority-vote based prioritization converges to the optimal value much faster than random allocation. Still, convergence is much faster for the toxicity dataset.}
\end{figure}

For our simulation analysis, we include one additional strategy: "Human Disagreement Prediction-Based Allocation", a randomly sampled human rater disagreement prediction label is used as the panel priority score. In general we can see that all strategies are slower to converge to the optimal value compared to the toxicity dataset (\Cref{fig:toxicity-simulation}). For example, under a Majority-vote Based panel allocation strategy, around 60\% of comments must be assigned to panel review to approximate the decision consistency benefits of universal panel review. In contrast, this was achieved in the toxicity dataset with just over 30\% of cases being assigned to panels. Still, model based panel allocation strictly outperforms random assignment. Thus even under strict conditions, our modeling approach is able to improve decision consistency and surface disagreements. 

\section{Think-Aloud Study: How is Venire Used in Practice?}

Given the positive results from initial qualitative and quantitative assessments, we decided to move forward with building and evaluating Venire. In this section we present the finalized Venire interface, and results of the final set of evaluation interviews. Given our findings around Venire's potential as an onboarding tool (\Cref{sec:prelim}), we had interview participants pretend to be newly added moderators to the r/ChangeMyView subreddit. We gave them a sandboxed moderation queue filled with real comments reported for violating r/ChangeMyView's Rule 2. Participants were then asked to think aloud while using Venire to make Rule 2 determinations and panel assignments. Broadly, our goal was to assess whether moderators would use the system as we intended, and to re-assess whether they viewed the system as valuable. Concretely, we sought to answer the following research questions: 

\begin{itemize}
    \item[] \textbf{RQ3a:} When do moderators consider flagging cases for panel review?
    \item[] \textbf{RQ3b:} How did users incorporate machine learning model recommendations into their decision-making?
    \item[] \textbf{RQ3c:} How do Venire's panel review system and model recommendations impact participants' experience learning a new community rule? 
    \item[] \textbf{RQ3d:} What kinds of communities do moderators foresee Venire working best in?
\end{itemize}

\subsection{Final Interface}

\begin{figure}[!t]
  \centering
    \fbox{\includegraphics[width=0.9\textwidth]{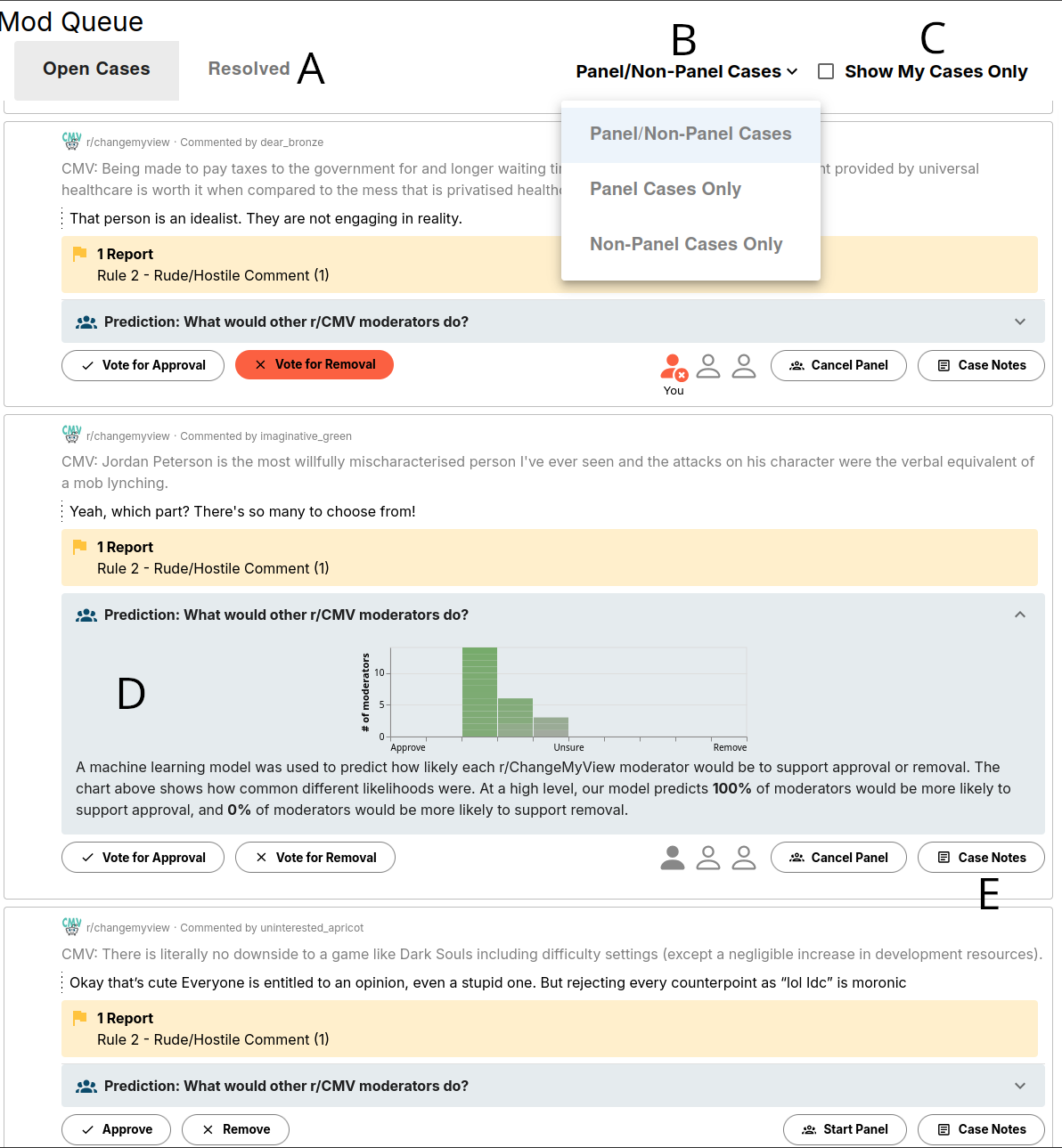}}
    \caption*{Labeled Venire interface. A: The button used to access Venire's log of resolved moderation cases. B: A filter that lets users look at only cases that are already in panel mode. C: A filter that lets users look at only cases they have voted on already. D: A collapsible visualization of model predictions for a specific case. Additional text is included here when cases are recommended for panel review. E: The button used to access a list of moderator-generated notes for a particular case. This can be used as a channel for case-specific deliberation between moderators.}
    \label{fig:main-ui}
  \hfill
\end{figure}

Based on the feedback from our initial interviews, we implemented the strict voting mock-up. We made a few updates to the interface based on lower-level feedback from our initial interviews. Most notably, we removed the "Review" modal and moved buttons directly onto the case card. This streamlined the workflow and reduced the number of necessary clicks. We also implemented the following features (\Cref{fig:main-ui}):

\begin{itemize}
    \item[] \textbf{Resolved Tab (A)}: We created a separate queue for past cases that have been reviewed by moderators. This was included because a few moderators (N=2) mentioned in the initial interviews that they were interested in reviewing the log of past panel decisions. (See \Cref{appendix:Venire} for an example of a resolved case)
    \item[] \textbf{Panel Filter (B)}: In the ``Resolved'' and ``Open Cases'' tabs, moderators are given the option to: see only cases in panel mode, see cases not in panel mode, or see both kinds of cases. This allows moderators to prioritize decision-making as they see fit.
    \item[] \textbf{My Cases Filter (C)}: We also gave moderators the ability to view only cases that they had previously ruled on. This includes cases in panel mode that they voted on, and cases not in panel mode for which they were the sole decision-maker. This allows moderators to check on the status of panels they previously participated in.
    \item[] \textbf{Panel Predictions (D)}: We updated our panel prediction from our initial mockup in \Cref{sec:prelim-prototypes}. Rather than dividing moderator predictions into binary categories of "Approve" and "Remove", we display a histogram of the model-produced prediction score for each moderator. To ensure that the prediction scores are interpretable, we calibrate the model using Platt scaling on a validation set. We also provide a text description, which contains an explicit, bolded recommendation for panel review in cases where the model predicts failure to achieve a supermajority amongst 70\% or more of the moderation team.
    \item[] \textbf{Case Notes (E)}: We provide moderators with a small chat interface which they can use to communicate their thoughts on specific cases, as this was requested a few times in our initial interviews (N=3). See \Cref{appendix:Venire} for the case notes interface.
    \item[] \textbf{Thread Context}: Clicking anywhere on a case card lets moderators view additional thread context for the reported comment. See \Cref{appendix:Venire} for the comment recreation interface.

\end{itemize}

Additionally, when moderators decline to start a panel, we occasionally display a pop-up recommending panel review (\Cref{appendix:Venire}). This happens under two conditions. First, if the model believes disagreements are broadly likely to occur for the case. We define this as when the model predicts moderators will not achieve a 70\% supermajority decision. Second, if the model predicts other moderators are likely to disagree with the decision \textit{inputted by the user}. We define this as when the model predicts 80\% of moderators or more will disagree with the inputted decision. This is similar to our \textbf{Predicted Disagreement+Majority Based Assignment} strategy from \cref{sec:technical}. Although we found that \textbf{Predicted Majority-Based Assignment} was more effective, we chose to include the disagreement score based alert as there may be qualitative differences that make these cases valuable to discuss.

\subsection{Recruiting and Protocol}

To recruit participants, we reached out to the moderators from the preliminary interviews. Five agreed to a follow-up interview. We recruited one additional moderator from P8's subreddit, giving us a total of 6 interviews. To train the final predictive model for Venire, we used the r/ChangeMyView, Rule 2 report dataset from our Prolific study (\Cref{sec:technical-prolific-protocol}). This time, however, we used the actual moderation decisions and moderator IDs from the r/ChangeMyView moderation log. In total, we used 3369 comments for training and 481 comments for validation and calibration. We populated the mod queue shown to participants with the 200 test set comments from the Prolific study. Our model achieved an accuracy of 75\% on this set of comments (AUROC: 0.84, Precision: 75\%, Recall: 79\%)

The interview consisted of three parts. First, we gave participants an overview of the system and informed them that they would be acting as a new r/ChangeMyView moderator. We then gave participants an overview of r/ChangeMyView and r/ChangeMyView's Rule 2, as in the Prolific study (\cref{sec:technical-prolific-protocol}). Participants were also given a guided tour of the interface. When describing the predictive model, we informed participants of the dataset the model was trained on, the input features the model used, and its accuracy. Next, participants were instructed to explore the interface and to think aloud while using the tool to make Rule 2 decisions, for around 20 minutes. The interviewer occasionally prompted participants to elaborate on why they made certain decisions. To improve ecological validity, the research team preset some of the mod-queue cases to be in panel mode or in the resolved queue at the beginning of the interview. This procedure is described in the \Cref{appendix:eval-presets}. Finally, after using the tool, moderators were asked a series of closing questions to reflect on their experience using Venire. Interviews lasted one hour long, and participants were compensated \$50 for participating.

\subsection{Results}

\subsubsection{When do moderators use panel review? (RQ3A)}

All moderators described starting panels when they were unsure of the best decision (N=6). Frequently, moderators described these cases as ``borderline.'' P2, for example, described starting panels when they could ``see an argument both ways.'' The precise reasons for why a case was considered borderline varied. Sometimes, moderators described it in terms of the severity of the infraction (e.g. ``This guy was sort of rude, but not really'' - P8). Other times, moderators referenced inherent ambiguity in the text. For instance, after reading a comment where one user called another user a Marxist, P2 said,  ``It does call into question if just saying that somebody is affiliated with a political party is an insult. Some people wouldn't mind that affiliation [...] but I know other people could take it just straight up as an insult.''

Occasionally, participants mentioned starting a panel when they felt a comment should be removed for reasons outside of Rule 2 (N=2). P3 described one instance of this, saying, ``I don't think [the comment] is contributing. So looking specifically at rule two, I think that it would be probably OK, but I'm sure that there's larger reason that we might want to remove it.'' Additionally, one participant, P7,  started a panel to preempt personal bias from influencing their decision, saying ``If I disagree with someone, I'm honestly more reluctant to remove their comment [...] , I would definitely move to a panel [...] especially since, at least for my subreddit, we tried to be as evenly political leaning as possible.''

\subsubsection{Incorporating Model Predictions into the Decision-making}

On the whole participants felt the model predictions did not sway their personal preference for removal or approval (N=3), but did influence their decision to start a panel (N=5). Participants differed slightly on how they factored in the model predictions into panel decisions. Most commonly, participants considered starting a panel when the model's predictions differed from their own decision (N=5). Less frequently, participants said they would start a panel when the model predicted a split in the mod team (N=2). This lined up surprisingly well with our findings around the empirical efficacy of majority-vote based and disagreement-based allocation strategies. As such, participants rarely saw the panel recommendation pop-ups. Interestingly, model predictions played a role in deciding when \textit{not} to start a panel (N=4). When the model predictions aligned with their removal decision, participants would decline to start a panel, usually to avoid creating extra work for others. Participants described using the model like this as a kind of ``gut check'' (N=4). Participants often described weighing their personal certainty alongside the model's confidence. P3 described this as follows:

\begin{quote}
    ``I would start a panel either if I was feeling unsure and the model was also unsure [...] Or if the model seemed to be quite sure of itself and I was disagreeing with the model. Whereas if I feel confident and the model is aligned, definitely I'm not going to start a panel. If I feel confident and the model is unsure... then I might start a panel? [...] And likewise if I felt unsure and the model seemed to be pretty certain, I probably wouldn't start a panel in that case. I would probably just go with what the model predicts''
\end{quote}

Participants were split on how often they looked at the model predictions. Most commonly, participants said they would not look at the prediction if they were confident in their decision (N=3). Some participants disagreed. For example, P7 said that they ``like to make sure, if I'm confident, that my confidence is well placed'' and that they usually checked the model predictions since ``it just takes a second and it don't hurt.'' A few participants tried to avoid biasing themselves by coming to their own decision before looking at the model prediction (N=2).


\subsubsection{Venire as an onboarding tool}  On the whole, participants found Venire to be beneficial to their experience as a ``new'' ChangeMyView moderator (N=5). Participants described the benefits of the panel system and the predictive model slightly differently. For the panel system, participants largely highlighted its benefits as a built-in channel for existing subreddit onboarding processes (N=3). Participants often contrasted the panel system with existing moderator group chats. P3 described the panel system as ``clean,'' since they ``don't want to start five different threads in the mod chat'' when they needed advice on multiple decisions.  P2 also felt that the panel system allowed them to be more active as a moderator since ``I can weigh in on cases that I have like a weaker opinion on, or that I'm not sure on" and that ``it also means that I can take an action on everything in the mod queue.'' 

In contrast, the predictive model was described as helpful for understanding subreddit norms, especially in places where the rules were not clearly specified (N=4). P9, for example, said that ``it made it so you can do a lot more on your own'' since ``you're not sitting there having to bother other mods asking a million questions.'' P2 echoed this sentiment saying ``different subreddits will have a different tolerance for what they consider uncivil'' and that the model predictions ``would give [them] confidence in understanding the quote-unquote culture of the subreddit.'' They summarized these sentiments neatly, describing the model as a ``a distilled version of past moderation decisions.'' 

\subsubsection{What kinds of communities should use Venire? (RQ3d)} Participants felt that Venire would be most effective on large subreddits (N=4). This was in part because those subreddits would have a ``larger training data set'' (P3), but also because in a large moderation team, moderators ``don't talk to each other very often.'' Participants also cited political subreddits (N=2) since they contained more "``gray area'' and ``nuance.'' Interestingly, a few moderators actually felt Venire would be more useful when the subreddit rules are poorly defined (N=2) since there would be more ``vague areas'' and ``ambiguity'' where the predictive model and panel system could help.

\section{Discussion}

\subsection{The Workload-Quality Tradeoff}

Venire is best understood as a system that empowers moderators to manage the tradeoff between maximizing decision quality and minimizing workload. Quantitatively, we measure decision quality by looking at how often moderation outcomes reflect the majority viewpoint amongst human raters. We measure workload by looking at how many human raters are involved in each decision. Venire's two features, the panel system and the ML model recommendation system, work in concert to help moderators balance these two concerns. The panel system gives moderators a formal way to incorporate more voices into a single decision, thereby improving decision consistency. The machine learning model, on the other hand, helps moderators identify which cases would most benefit from panel review, minimizing workload. The key intuition behind the system is that many moderation cases are straightforward for moderators, and would not benefit from panel review.

Still, our quantitative evaluations in \Cref{sec:technical} belie more nuanced mechanisms through which Venire can impact decision quality and workload. With respect to decision quality, we note that the direct benefits to decision consistency reflect only a fraction of the greater value of surfacing disagreements. In our interviews, we found that moderation teams already discuss discrepancies and update policy to iron out disagreements. Ideally, Venire can help moderators identify points of disagreement more quickly, resulting in more policy updates that reduce the need for future panel review. 

With respect to workload, our interviews point to additional benefits to Venire we had not initially anticipated. In our preliminary interviews, moderators speculated that Venire could act as an onboarding tool for new moderators. Moderators felt that ML-guided panel reviews could act as a ``safety net'' for new moderators, minimizing the risk that a new moderator will make a decision out of alignment with the moderation team as a whole. Crucially, if Venire makes it easier to recruit new moderators, it could actually lead to a decrease in per-person moderation workload. Our evaluation interviews lend credence to this: moderators described the ML model recommendations as a ``distilled'' version of a subreddit's moderation decision history. Moderators found this helpful for understanding the norms of a particular subreddit's moderation style, especially in places where the rules were not fully specified. Moderators also noted that the panel review system allowed them to ``take an action on everything in the mod queue'' rather than ``leaving the complicated stuff for someone else.'' In the status quo Reddit moderation queue, moderators may spend time reviewing a case but defer on making a decision, contributing to group inefficiency. Panel review can reduce overall workload if it encourages moderators to weigh in early, and prevents cases from sitting in the queue for an extended duration. 

\subsection{Future Considerations Before Deploying Venire}

While we believe our work represents a comprehensive investigation into the feasibility of Venire, in this section we highlight a few factors that should be explored more closely before deploying Venire in the wild. 

\subsubsection{Training Venire's prediction model} In an ideal world, a Venire instance could be trained for a particular community using only existing moderation log data. That way, moderators would not need to spend time labeling additional training data points. However, real-world moderation log data is largely singly labeled. That is, each case has a single label associated with it, supplied by a single decision-maker. In our technical evaluations, we found that panel review was significantly less efficient when assigned by a model trained on singly labeled data. Although we did not disentangle the effects of a singly labeled training dataset from other sources of performance dropoff (e.g. the dataset containing fewer comments, different task conditions, etc.), this concern is worth further investigation. Notably, the panel system itself naturally produces multiply-labeled data. Thus, it is possible that Venire could see performance improvements if the predictive model is retrained after a sufficient number of panel decisions have been made. The predictive model could even be turned off until enough human-flagged panel decisions have been made. 

Moderation log data differs from our crowdsourced datasets in another crucial way. In the actual moderation queue, moderators \textit{choose} which cases they want to make a decision on. This is in contrast to the crowdsourced datasets, where workers were randomly assigned to cases. It is possible that selection bias effects present in the real moderation log data could degrade the quality of model predictions. When we trained our final prediction model on real r/ChangeMyView log data, we evaluated the model's ability to make predictions on a hold out portion dataset. However, if strong selection effects are at play, this test set cannot capture how well the model predicts every moderator's decision on every case. Constructing such a test set would require asking moderators to do additional labeling work outside the moderation queue. Future work should explore the impact of selection effects on Venire to more accurately characterize its expected performance. 

\subsubsection{Effects on moderator decision-making} Venire's panel system and ML model recommendation system may have unintended effects on moderators' decision-making processes. Understanding and mitigating these effects is crucial to successfully deploying Venire in a real online community. Our final set of interviews with moderators point to a few promising areas for further investigation. With respect to the panel system, one concern is that the voting-based aggregation could \textit{decrease} the amount of discussion that moderation teams are having. Participants in our final interviews often described the panel system as ``more organized'' or ``more efficient'' than consulting a moderator group chat. While this was framed as a positive factor, our interface may not support robust discussion in the way that a group chat does. Adding stronger support for deliberation or integrating with chat software like Discord could address this. One participant, P3, suggested adding a deliberation phase to the panel decision-making process after the votes have been cast. 
\begin{quote}
    ``If I felt very strongly one way and I was the minority [..] I would want to be able to hop on discord with the other two moderators and say I feel very strongly, [...] I'd want to have a larger discussion. Because obviously, if something gets approved, it can always be removed later. It's harder to go in the other direction.''
\end{quote}

 Venire's ML model recommendations could also negatively impact moderator decision-making. The human-AI interaction literature contains many examples of how an AI model recommendation can bias human decision-making. One key concern is that moderators may defer to the model instead of relying on their own judgment. This could lead to a kind of AI information cascade \cite{anderson1997information} where each moderator is overly influenced by the model prediction, and does not communicate their personal judgment to other decision-makers. Generally, our interview findings suggest that this is not the case. Moderators described making up their own minds before looking at the ML model recommendations. Rather than influencing their decision to remove or approve a comment, participants instead described the model as playing a role in their choice to start a panel or not. Most commonly, moderators described either starting a panel when the model disagreed with them, or being convinced not to start a panel when the model agreed with them. This indicates that the system is working as intended: a signal of the individual moderator's judgment is still being transmitted, regardless of whether a panel is started or not. Still, our interviews rely on moderators to self-report their thought processes---we cannot rule out the possibility that moderators are biased without realizing it.  

Echoing recommendations from \citet{hullman2024decision}, we suggest that future research on the biasing effects of Venire formally model the underlying decision problem. Most importantly, researchers should consider carefully when a decision-maker ought to seek panel review. This will allow us to identify cases where cognitive biases lead to suboptimal outcomes. To guide future work, we highlight two important aspects of our content moderation task that a formal model should reflect. First, our content moderation task is purely subjective---the predictive model is only correct insofar as it is able to anticipate different decision-makers' preferences. This is in contrast to other human-AI decision-making tasks, where there is an external ground truth that the human and the model are trying to predict. Second, the predictive model has access to the same information as the decision-maker. In short, the AI model cannot provide a human rater with a signal into their personal preference for a case. Thus, we suggest that a rational decision-maker should only use the model as a signal of how other decision-makers will act, and not to inform their own independent judgment. A more formal model of the problem can be used to test this assumption, and better capture places where an AI model can distort moderation outcomes. 

\subsubsection{Identifying patterns of disagreement} Moderators were interested in reviewing Venire's log of panel decisions to refine moderation policy. This process aligns with what Cullen and Kairam refer to as ``reflective practice'' in moderation \cite{cullen2022reflexive}. Future work could augment Venire's interface to support reflective practice more explicitly. One way to do this might be to help moderators identify \textit{patterns} of disagreement. This could be done at the interface-level by adding support for tagging and grouping panel cases within the moderation log. Technical approaches such as clustering techniques or large language models could also be used to summarize the log of disagreements automatically. Model interpretability techniques that provide explanations for specific predictions of disagreement could facilitate this process. For instance, influence function-based explanations \cite{koh2017understanding} identify the points in the training dataset that inform a specific model prediction. When the model predicts disagreement, influence functions could be used to identify historical cases in the moderation log which were ruled inconsistently between moderators.  
\section{Conclusion}

In this work, we present a comprehensive exploration of Venire, a machine learning-guided panel review system for community content moderation. Through a series of three studies, we provide quantitative and qualitative evidence that Venire improves decision consistency between moderators and surfaces latent disagreements within moderation teams. More broadly, we argue that Venire helps moderators navigate the tradeoff between maximizing decision quality and minimizing workload. Unlike prior work on AI content moderation systems, Venire represents an attempt to use machine learning to more efficiently allocate human decision-makers, rather than replace them outright. We call for more CSCW research that supports reflective practice amongst moderators, empowering them to refine community policy through case-based reasoning.   

\bibliographystyle{ACM-Reference-Format}
\bibliography{sample-base}

\appendix
\section{Preliminary Interview: Moderator Recruitment Details}\label{appendix:prelim-recruitment}

Several measures were taken to avoid spamming moderators with recruitment messages. First, we avoided subreddits that we had recently recruited for other studies. Second, we messaged subreddit's such that no moderator received a recruiting message twice. To do this, we constructed a network of our filtered communities where each edge corresponded to a community that shared at least one moderator. We then ran an independent set solver to find a subset of communities that did not share any moderators. We messaged only these communities. Recruiting messages were sent to five communities a day until the desired number of participants was reached. 

\section{Prolific Study: Survey Materials}\label{appendix:prolific-materials}

\subsection{Content Warning}

In this task, you will see comments that were previously posted to r/ChangeMyView (r/CMV), some of which were removed by the community's moderators. The research team conducting this study feels that a few of these comments are potentially offensive. If you do not feel comfortable viewing such text, we advise you to exit the study. Otherwise, you can press the next button to continue on to the next page. You may exit the study at any point in time, though you will not receive payment unless you complete the full study.

Specific triggers you might encounter include references to:
\begin{itemize}
    \item Racism
    \item Misogyny
    \item Sexual assualt
    \item Gun violence
    \item Homophobia/Transphobia
    \item Body shaming
    \item Anti-semitism
    \item Ableism
\end{itemize}

\subsection{Introduction to r/ChangeMyView}

\noindent \textbf{Page 1 Instructions}: ``In this task, you'll be asked to pretend to act as a content moderator for r/ChangeMyView, a Reddit community. On r/ChangeMyView, people post about a viewpoint they hold that they want to have changed. In the comments section, other users submit arguments to change the original poster's (OP) view. Some posters' viewpoints may be controversial or offensive -- still, such posts are allowed as long as the poster is open to having their mind changed. Below is an example of a post on r/ChangeMyView:''

\noindent \textbf{Post title}: ``CMV:Taylor Swift is an average musician''

\noindent \textbf{Post body}: `` I have seen many posts and heard people say many things to hype up Taylor Swift. They say Taylor Swift is a better vocalist than Adele. They say Taylor Swift is a better performer than Beyonce. I even heard someone say that Taylor Swift is one of the best songwriters of all time(when people like Alicia Keys and Bruno Mars exist). I don’t think Taylor Swift is a terrible artist. She can actually hold a tune unlike Jennifer Lopez or Selena Gomez. Her Performances aren’t as high energy and powerful as a Beyonce performance but something I do appreciate is that Taylor Swift can play instruments while singing which is something not many performers can do. However I don’t think Taylor Swift is anywhere close to Beyonce when it comes to performing. Something I do appreciate about Taylor Swift is that she story tells through her music however all of her music is a breakup story. Where is the variety in that? 

I understand that Taylor Swift is one of the biggest artists out right now but in my opinion I don’t think Taylor Swift is as talented as people make her sound. She is average in all aspects of music. Nothing about her screams “I’m the best at what I do”. Nothing about her stands out among the crowd of much better musicians. ''

\noindent \textbf{Page 2 Instructions}: ``If a commenter changes the OP's mind, the OP can give them a 'delta' as a reward. Deltas act as a point system on r/ChangeMyView. The community maintains a leaderboard to show which users have earned the most deltas over time. Below is an example of a reply comment and a delta award. The "OP" symbol next to the second reply shows it's from the original poster. The OP types '!delta' into their message (typing $\delta$ also works)  to award a delta at the bottom of the second reply. You may occasionally see references to the delta system when labeling comments for this task.''

\noindent \textbf{Example parent comment}: `` She been touring for the past year (and will be for almost another year) in all different time zones each week, performing 3.5 hours of her discography. Not only is she doing dancing through those entire songs, but she also has elaborate sets and performances throughout. Not only that, but she has full songs playing the guitar and piano on top of that. These are sold out arenas around the world

And within that crazy tour, she’s continuing to write, produce and put out new music. And when we say.l new music, it’s a double album consisting of 30 tracks. Then she records music videos. She has 11 studio albums and averaging like 20+ songs each. Thats simply incredibly dedication of an artist to give their entire self to their craft.

She’s also really supportive to other artists - she’s the only person you’ll see at an award show standing at every performance. Generally unproblematic and makes friends with everyone. Invites fans all the time to events just from her or her team seeing their dedication on social media/etc. I mean she even invited fans to her house once. Every show and event you’ll see her having invited hand selected people from her fan base.

I understand her 'generic white girl pop music' might not be for everyone but girl credit where it’s due that she put the damn work in to be where she is now.''

\noindent \textbf{Example response from OP}: ``Yea someone else mentioned that vocally and performance wise she is average when compared to other well known singers. But they said her personality is what makes her stand out. I can appreciate that. I don’t really like Taylor’s music like that but she never causes any drama online, she always supports smaller artists and many artists already say that Taylor is a huge inspiration to them. Even though Taylor has had many breakups she never takes these breakups online like many other celebrities. She keeps her life confidential which I like. !delta ''

\subsection{Rule 2 Overview}

r/CMV has a few rules for commenting. In this task you'll review real r/CMV comments and judge whether they break CMV's Rule 2, which states: 'Don't be rude or hostile to other users.' Below are some basic guidelines to help you decide if Rule 2 is violated. You can view these guidelines on later pages.

\begin{itemize}
    \item Rule 2 only forbids hostility towards other r/CMV users:Rule 2 bans comments that are rude or hostile to other users (e.g., ``You're an idiot''). It doesn't apply to comments containing broadly offensive viewpoints, or insults toward public figures, institutions, or general categories of people. Although such comments can be offensive, r/CMV often aims to persuade people to change these views. An exception is if a group insult is clearly aimed at another commenter. For example, saying ``People who wear glasses are stupid'', after another commentor says they wear glasses. It's more acceptable to criticize arguments than people -- e.g., ``That argument is bad'' is okay, but ``You are bad'' is not. Still, there are limits. Calling an argument ``moronic'' is probably still considered rude or hostile.

    \item Excessive passive-aggression is considered hostile: Rule 2 applies to comments that are excessively passive-aggressive, even if they do not contain an overt insult. A common example is claiming that another user "lacks reading comprehension abilities".

    \item Rule 2 applies if any part of a comment contains rudeness or hostility. If even one sentence in a three paragraph comment is hostile, the comment can be considered for removal.

    \item There is some subjectivity that goes into making a rule 2 determination. Please use your best judgment---although we will do some quality checks on your answers, we recognize that not everyone will interpret a comment the exact same way

\end{itemize}

\subsection{Practice Comments}
\noindent \textbf{Comment 1}: ``you are an idiot. any BEING that can feel pain is not object. Simple as that.''

\noindent \textbf{Comment 1 Explanation (User Says Rule 2 Applies)}: ``Good job! This comment was removed by the r/ChangeMyView moderators. Calling another user an 'idiot' is a clear example of rudeness''

\noindent \textbf{Comment 1 Explanation (User Says Rule 2 Does Not Apply)}: `` Unfortunately, the r/ChangeMyView moderators removed this comment. Calling another user an 'idiot' is a clear example of rudeness. Please select 'The comment violates Rule 2' to continue''

\noindent \textbf{Comment 2}: ``I have a Juris Doctor degree. I am very comfrey in my understanding of the term 'rights'. The arguments you are holding onto can at best be described as purely semantic, which means I cannot help you or so vapidly stupid I'm honestly surprised you managed to create a Reddit account.''

\noindent \textbf{Comment 2 Explanation (User Says Rule 2 Applies)}: ``Good job! This comment was removed by the r/ChangeMyView moderators. Although a bit less straightforward than the previous case, the commentor is insulting someone's intelligence by indicating surpise at their ability to create a Reddit account''

\noindent \textbf{Comment 2 Explanation (User Says Rule 2 Does Not Apply)}: ``Unfortunately, the r/ChangeMyView moderators removed this comment. Although a bit less straightforward than the previous case, the commentor is insulting someone's intelligence by indicating surprise at their ability to create a Reddit account. Please select 'The comment violates Rule 2' to continue''

\noindent \textbf{Comment 3}: ``All of those statements are pretty misogynistic. Women don't \*need\* to do anything they don't want to in their dating life, and frankly their strategies for pursuing men are their own business. Dating isn't, and shouldn't be, about fairness and equality. Romantic engagements are deeply personal and people are allowed to like or dislike whatever they want in a person. So, yeah, this kinda proves how gross the incel mindset is.''

\noindent \textbf{Comment 3 Explanation (User Says Rule 2 Applies)}: `` Unfortunately, the r/ChangeMyView moderators did not remove this comment. This case is borderline. Although one could argue the term 'misogynist' indicates hostility, the commentor is using the label to refer to another user's arguments, rather than to the user themselves. Please select 'The comment does not violate Rule 2" to continue' ''

\noindent \textbf{Comment 3 Explanation (User Says Rule 2 Does Not Apply)}: `` Good job! This comment was not removed by the r/ChangeMyView moderators. This case is borderline. Although one could argue the term 'misogynist' indicates hostility, the commentor is using the label to refer to another user's arguments, rather than to the user themselves.''

\subsection{Disagreement Prediction Instructions}

\noindent Thank you for completing the previous block of comments. 

\noindent In the next section, you'll review another 20 comments. Each comment in this section will also be reviewed by 6 other Prolific workers.

\noindent You will be asked to do two things for each comment:
\begin{enumerate}
    \item You will be asked to determine whether the comment violates Rule 2.
    \item  You will be asked to predict whether the other 6 raters will be be in high consensus or low consensus over the violation of Rule 2.

\end{enumerate}

\noindent We say there is high consensus amongst the other 6 raters, if one or zero of the other raters dissent from the majority viewpoint (i.e. a 6/0 or 5/1 split). We say there is low consensus amongst the other raters if the other raters are evenly split, or if two raters dissent from the majority viewpoint (i.e. a 3/3 or 4/2 split).

\noindent You'll get a 10-cent bonus every time you correctly predict the consensus level for a comment -- up to \$2.00 for a perfect score. Bonuses will be given within a week.

\noindent \textbf{Question text}: ``6 other raters will be shown this comment. Consider how they will respond to the previous question. Will there be high consensus amongst these raters over whether 2 has been violated (e.g. 6/0 or 5/1 split), or low consensus? (e.g. 3/3, 2/4, or 4/2 split)''

\noindent \textbf{Answer Choices }: ``High Consensus'', ``Low Consensus''
\newpage
\section{Venire Interface}\label{appendix:Venire}

\begin{figure}[!h]
  \centering
    \fbox{\includegraphics[width=0.9\textwidth]{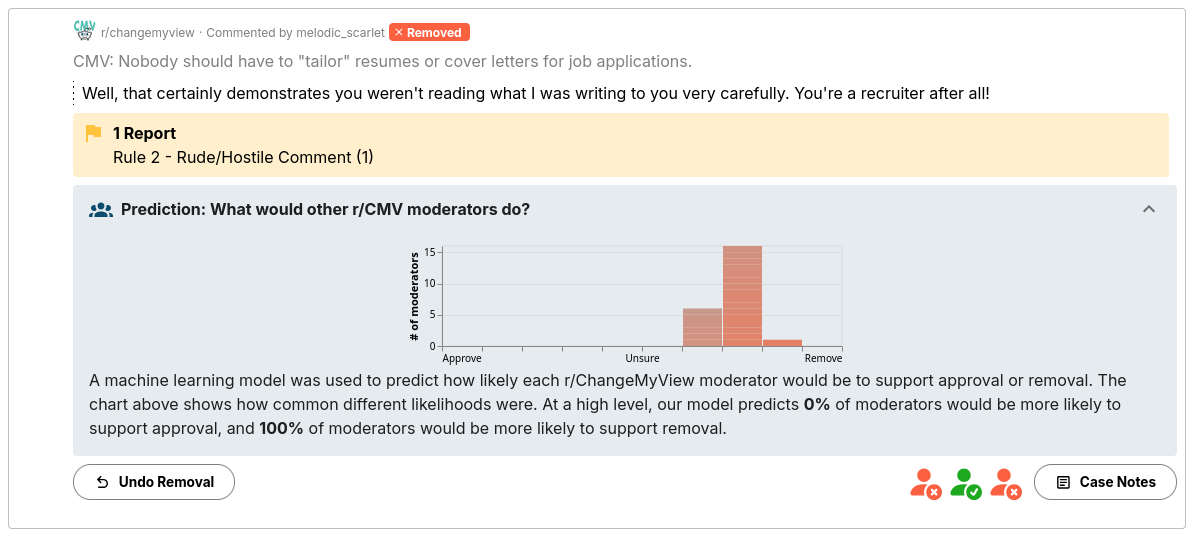}}
    \caption*{An example of a previously handled moderation case in Venire's resolved queue. Moderators can see the final status of the comment, and how the panel of decision makers voted} \label{fig:resolved-ui}
\end{figure}

\begin{figure}[!h]
  \centering
    \fbox{\includegraphics[width=0.4\textwidth]{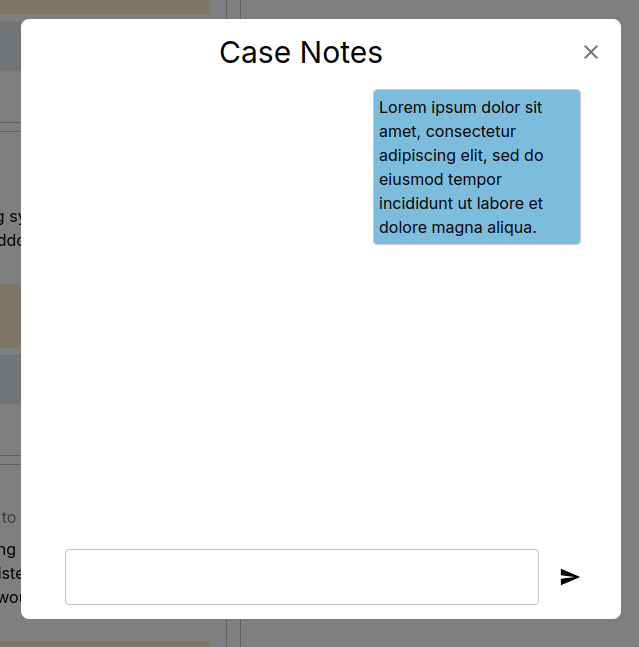}}
    \caption*{The Venire case notes interface. Moderators can this chat-like feature to handle case-specific deliberations}\label{fig:ui-case-notes}
\end{figure}

\begin{figure}[!h]
  \centering
    \fbox{\includegraphics[width=0.6\textwidth]{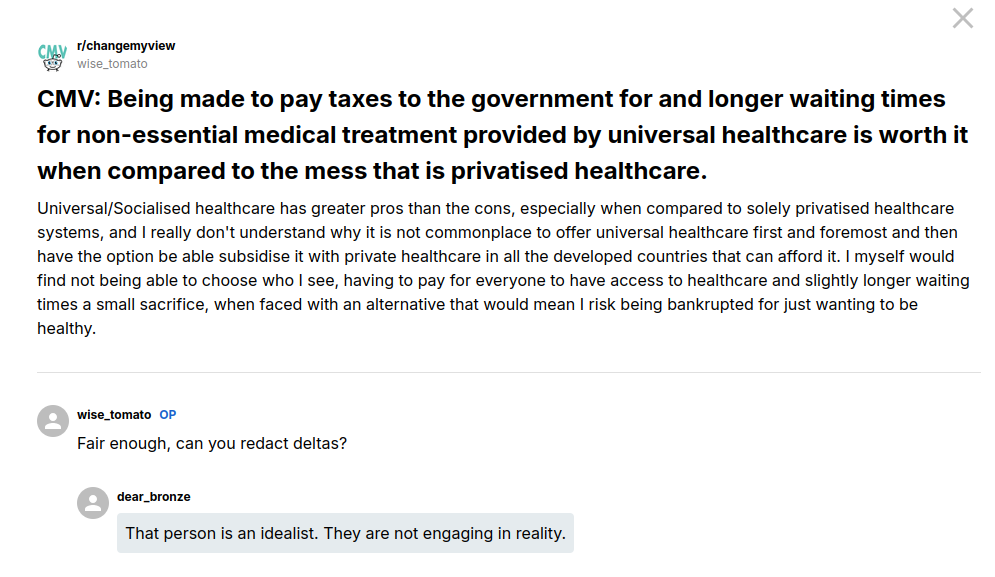}}
    \caption*{Interface recreation of additional thread context. Users can access the thread context for a case by clicking anywhere on the case card} \label{fig:ui-sidebar-context}
\end{figure}

\begin{figure}[!h]
  \centering
    \fbox{\includegraphics[width=0.7\textwidth]{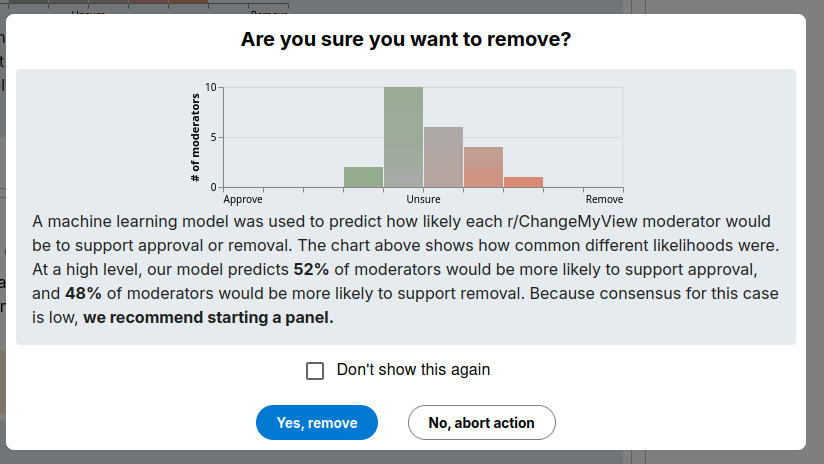}}
    \caption*{Pop up alert recommending the user pursue panel review. The pop-up interface includes the visualization and text from the panel prediction tab} \label{fig:ui-pop}
\end{figure}

\newpage
\section{Evaluation Interview: Presetting Moderation Queue}\label{appendix:eval-presets}

To improve ecological validity of the evaluation interviews, the research team preset some of the mod-queue cases to be in panel mode or in the resolved queue at the beginning of the interview. This was done using randomly sampled labels provided by Prolific raters. We set 13 cases to panel mode in places where a Prolific rater had predicted low consensus. 3 of these went into the resolved queue and 10 went into the open cases queue. Another 3 non-panel cases were included in the resolved queue. A total of 25 votes and Rule 2 decisions were preset using the Prolific ratings

\end{document}